\documentclass[
aps,nofootinbib,
showpacs,showkeys,preprint
tightenlines,preprintnumbers,] {revtex4}

\usepackage{epsf,epsfig,subfigure,graphicx,amsmath,amssymb 
}
\usepackage{color}
\newcommand{\dis}[1]{\begin{equation}\begin{split}#1\end{split}\end{equation}}
\newcommand{\ie}{{\it i.e.~}}

\newcommand{\Qanom}{Q_{\rm anom}}
\newcommand{\Qem}{Q_{\rm em}}

\newcommand{\cagg}{c_{a\gamma\gamma}}

\newcommand{\gev}{\,\textrm{GeV}}

\newcommand{\eV}{\,\mathrm{eV}}
\newcommand{\Mp}{M_{\rm P}}

\newcommand{\Mv}{{M_{\rm vec}}}
\newcommand{\Mvt}{{$M_{\rm vec}$}}
\newcommand{\Mg}{{M_{\rm GUT}}}
\newcommand{\Mgt}{$M_{\rm GUT}$} 
\newcommand{\Mi}{M_{\rm int}}
\newcommand{\Mit}{$M_{\rm int}$}
\newcommand{\vew}{$v_{\rm ew}$} 

\newcommand{\Uanom}{U(1)$_{\rm anom}$}

\newcommand{\UPQ}{U(1)$_{\rm PQ}$}
\newcommand{\SUflip}{SU(5)$\times$U(1)$_X$}
 
\def\sw0{{$\sin^2\theta_W^0$}}

\newcommand{\Z}{{\bf Z}}

\def\smg{{SU(3)$_C\times
$SU(2)$_L\times$U(1)$_Y$}}
\def\smfa{{SU(3)$_C\times
$SU(2)$_L\times$U(1)$_Y\times$U(1)$_{\rm anom}$}}
 
\def\E6{{\rm E_6}}

\def\EE8{{\rm E_8\times E_8'}}

\def\one{{\bf 1}}

\def\two{{\bf 2}}
\def\five{{\bf 5}}
\def\ten{{\bf 10}}
\def\tenb{\overline{\bf 10}}

\def\fiveb{\overline{\bf 5}}

\begin{document}

\draft

\title{The anomalous \Uanom~ symmetry and flavors from an SU(5)$\times$SU(5)$'$ GUT in $\Z_{12-I}$ orbifold compactification} 
  
\author{Jihn E.  Kim}
\address
{Department of Physics, Kyung Hee University, 26 Gyungheedaero, Dongdaemun-Gu, Seoul 02447, Republic of Korea, and\\
Center for Axion and Precision Physics Research (IBS),
  291 Daehakro, Yuseong-Gu, Daejeon 34141, Republic of Korea
}

\author{Bumseok Kyae}
\address
{Department of Physics, Pusan National University, 2 Busandaehakro-63-Gil, Geumjeong-Gu, Busan 46241, Republic of Korea}

\author{Soonkeon Nam}
\address
{Department of Physics, Kyung Hee University, 26 Gyungheedaero, Dongdaemun-Gu, Seoul 02447, Republic of Korea}

\begin{abstract} 
In string compactifications, frequently there appears the anomalous U(1) gauge symmetry which belonged to E$_8\times$E$_8'$ of the heterotic string. This anomalous U(1) gauge boson obtains mass at the compactification scale ($\approx 10^{18\,}\gev$) by absorbing one pseudoscalar (corresponding to the model-independent axion) from the second rank anti-symmetric tensor field $B_{MN}$. Below the compactification scale, there results a global symmetry \Uanom~whose charge $\Qanom$ is the original gauge U(1) charge. This is the most natural global symmetry, realizing the ``invisible'' axion. This global symmetry \Uanom~is suitable for a flavor symmetry. In the simplest  compactification model with the flipped SU(5) grand unification, we calculate all the low energy parameters in terms of the vacuum expectation values of the standard model singlets.
 
\end{abstract}

\maketitle

\section{Introduction}
\label{sec:Intro}

Low energy global symmetries in the effective theory are of fundamental importance in the strong CP solutions \cite{PQ77} and cosmology \cite{Preskill83}. From the bottom-up approach, the Kim-Shifman-Vainstein-Zhakarov (KSVZ) axion model \cite{Kim79,Shifman80} and the Dine-Fischer-Srednicki-Zhitnitsky (DFSZ) axion model  \cite{Dine81,Zhit80} are of practical interests.\footnote{For the DFSZ model, only in the supersymmetric extension of the standard model the fine-tuning problem  \cite{Dreiner14} is evaded by the $\mu$ term \cite{KimNilles84}.} However, these global symmetries might be badly broken by gravitational effects \cite{Barr92}.

On the other hand, a consistent top-down approach, the so-called string models, is not allowing any global symmetry. In compactifications of the heterotic string \cite{Gross84}, there always exists the pseudoscalar degree from the second rank antisymmetric tensor field $B_{\mu\nu}\,(\mu,\nu=1,2,3,4)$ \cite{GS84}, which is the so-called  ``model-independent axion (MI-axion)'' \cite{Witten84}. If the MI-axion is physical, its decay constant is of order $10^{15\,}\gev$ \cite{ChoiKim85} which is marked in Fig. 1 of Ref. \cite{KimPLB16}. If this MI-axion degree is removed at the compactification scale, a global U(1) symmetry can survive down to realize the ``invisible'' axions at the intermediate scale \Mit~ \cite{Kim79,Shifman80,
Dine81,Zhit80}. This happens in compactifications with an anomalous U(1) gauge symmetry \cite{Anom87}.  The anomalous U(1) gauge symmetry is a U(1) subgroup of  $\EE8$ gauge group, and the corresponding gauge boson obtains mass at the compactification scale($\approx 10^{18\,}\gev$) by absorbing the MI-axion degree. In this case, a global symmetry   called \Uanom~is surviving down to a lower energy scale. Note that the so-called ``model-dependent  axions (MD-axions)'' from $B_{MN} \,(M,N=5,\cdots,10)$ \cite{Witten85} do not match to any U(1) subgroup of  $\EE8$ because the heterotic string has only one anomalous U(1) gauge symmetry. Because there is no global symmetry except the \Uanom~global symmetry, the MD-axions must be removed at the compactification scale unless they become accidentally light \cite{Wen86}. So, those used in Refs. \cite{Kim00} must be accidentally realized. In string compactification, accidental symmetries are pointed out to be related to axions \cite{ChoiAcc,Accions} and $R$ symmetry \cite{Nilles09}.
Here, we identify \Uanom~as the needed Peccei-Quinn (PQ) symmetry \cite{PQ77} for the ``invisible'' axion \cite{Kim79,Shifman80,
Dine81,Zhit80}.

 This leads us to consider a reasonable compactification model with a \Uanom~symmetry in a full detail. In this paper, we choose the model presented in Ref. \cite{Huh09}, 
 based on $\Z_{12-I}$ orbifold compactification.\footnote{Recently, a comparison of orbifold compactifications and free fermionic models has been studied \cite{Athanasopoulos16}.}
  We could have chosen the model presented in Ref. \cite{JHKim07} also, which however contains much more singlets and hence is more complicated to be presented here completely. Even though we present the analyses in the specific model, the current method can be applied to any model toward obtaining a complete knowledge on the ``invisible'' axion and flavor parameters.

The low energy gauge group obtained in \cite{Huh09} is \SUflip$\times$U(1)$^6\times
$SU(5)$'\times$SU(2)$'$ where the primed non-Abelian groups are from the hidden sector E$_8'$. The first two factors \SUflip~is the so-called rank-5 flipped-SU(5) \cite{Barr82,DKN84,Ellis89}. Being a GUT, the flipped-SU(5) must resolve the doublet-triplet splitting problem in the Higgs quintets $\five$ and $\fiveb$: ``Why are the color triplets superheavy while Higgs doublets remain light?'' In this paper, we show how the splitting is realized in terms of the complete spectra in the model. In fact, we go beyond the dimensional analyses.

The global symmetry \Uanom~is beyond the flipped-SU(5) and hence it can be used as a family symmetry. Since we know all the quantum numbers $\Qanom$ of the global symmetry \Uanom, we can obtain the order of magnitudes of all the Yukawa couplings, resolving the family parameters, \ie we can obtain mass matrices of the SM fermions. Basically, it turns out that the flavor matrices are given by the multiples of Yukawa coupling constants \cite{Nilles09} instead of the mass power suppressions via the Froggatt-Nielsen powers \cite{FN79}.

In Sec. \ref{sec:Charges}, we present the definition of quantum numbers and express $\Qanom$ in terms of six U(1) gauge charges of  $\EE8$ and derive which pair of $\five$ and $\fiveb$ are remaining toward the needed pair in the SUSY SM. Here, we also discuss the 't Hooft mechanism which is working for transferring the global symmetry down to the axion window.
In Sec. \ref{sec:MassScale}, we discuss mass scales in the model where \Uanom~is surviving as a PQ symmetry down to an intermediate scale. In Sec. \ref{sec:Yukawa}, we present the Yukawa mass matrices of $\Qem=+\frac23$ and $-\frac13$ quarks, $\Qem=-1$ charged leptons, and light SM neutrinos.
Section \ref{sec:Concl} is a conclusion. In Appendix, the 't Hooft mechanism in the compactification process is discussed.  For this occasion, we  present correct entries for the previous Tables of Ref. \cite{KimPLB14}, taking into account its erratum. 

\section{Global charges and one pair of Higgs doublets in SUSY}
\label{sec:Charges}

In the open string theory with $n$ Chan-Paton factors, string amplitudes are U($n$) invariant for example. This U($n$), constructed with $n$(fundamental)--$\bar{n}$(anti-fundamental) is the world sheet global symmetry, viz. p.374 of Ref. \cite{CK06}. Now in the target space, this is coordinate ($x_M$) dependent and hence  U($n$) is promoted to a gauge symmetry, which is the reason that string theory does not allow any global symmetry. The basic reason might be the string movement in the world sheet. However, if location of the string is fixed at a fixed point, variation of the string in the world sheet is not allowed and hence in the target space global symmetry may not be promoted to a gauge symmetry.  Fixed points are present in the symmetric orbifold compactifications, and the existence of a global symmetry is not ruled out. But, in the smooth compactifications there will be no global symmetry. Thus, an anomalous U(1) may not arise in smooth compactifications. To obtain a global symmetry producing an anomalous U(1) and hence the ``invisible'' axion, let us consider the orbifold compactification. 

 An  $\EE8$ heterotic string model compactified on $\Z_{12-I}$ orbifold gives  SU(5)$\times$SU(5)$'\times$SU(2)$'$ with seven U(1)'s \cite{Huh09}.  This GUT model has been studied for various aspects in Refs. \cite{KimPLB13, KimPLB14,KimNam16}.
 Extra U(1)'s  are just a problem in orbifold compactification. In Calabi-Yau compactifications, for example, the rank due to extra U(1)'s is easily reduced. But there is one important U(1) factor which is a part of the flipped SU(5) GUT \cite{Barr82}. Because of the difficulty of obtaining an adjoint representation for a Higgs multiplet for breaking SU(5), the flipped SU(5) is  
  probably the most favorable GUT in the orbifold compactification \cite{CK06}.\footnote{The Pati-Salam model \cite{PSgut73} is also good, but to break the part SU(2)$_L\times$SU(2)$_R$, contained in the PS model, down to SU(2)$\times$U(1), one needs a VEV of $\Delta\,(\equiv{\bf 3}$ under SU(2)$_R$), hence it is like an adjoint of SU(2). Anyway, because of this complexity, we may say that  the flipped SU(5) is `most favored' from string compactification.
  }
    So, firstly we pay attention to the factor  SU(5)$_{\rm flip}$  where our definition of  SU(5)$_{\rm flip}$ is containing a gauge group U(1): \SUflip. The second is the anomalous U(1). Except these two U(1) factors, U(1)$_X$ and \Uanom, the non-Abelian gauge group is SU(5)$\times$SU(5)$'\times$SU(2)$'$, and the rest anomaly free factors are $\tilde{\rm U}(1)^5$.
Since the rank of the original gauge group  $\EE8$ is 16, the total number of U(1) factors is 7. Let us name their charges as $X$ and $Q_i\,(i=1,\cdots,6)$. $\Qanom$ is a linear combination of  $Q_i\,(i=1,\cdots,6)$. In Ref. \cite{KimPLB14}, these charges are defined on the lattice as,\footnote{For the definition, see, Ref. \cite{CK06,KK07}.}
\begin{eqnarray}
&&  X=(-2,-2,-2,-2,-2\,;\,0^3)(\,0^8)'  ,\label{eq:X}\\[0.5em]
&&   \Qanom= 84Q_1+147Q_2 -42Q_3-63Q_5- 9Q_6,\label{eq:Qanom}
\end{eqnarray}
where
\begin{eqnarray}
 Q_1&=&(0^5;12,0,0)(0^8)',\nonumber \\[0.3em]
 Q_2&=&(0^5;0,12,0)(0^8)',\nonumber\\[0.3em]
 Q_3&=&(0^5;0,0,12)(0^8)',\nonumber\\[0.3em]
 Q_4&=&(0^8)(0^4,0;12,-12,0)',\nonumber\\[0.3em]
 Q_5&=&(0^8)(0^4,0;-6,-6,12)',\nonumber\\[0.3em]
Q_6&=& (0^8)(-6,-6,-6,-6,18;0,0,6)'. \nonumber
\end{eqnarray}

In the orbifold compactification, frequently there appears an anomalous U(1)$_A$ gauge fields $ {A}_\mu$ from a subgroup of $\EE8$ \cite{Anom87}. The charge of this anomalous  U(1)$_A$ is given in Eq. (\ref{eq:Qanom}).  In addition, the anomaly cancellation in 10 dimensions (10D) requires the so-called Green-Schwarz (GS) term in terms of the second rank antisymmetric-tensor field $B_{MN}\,(M,N=1,2,\cdots,10)$ \cite{GS84}. This GS term always introduces the  MI-axion  $a_{\rm MI}$, $\partial_\mu a_{\rm MI}\propto \epsilon_{\mu\nu\rho\sigma}
H^{\nu\rho\sigma}\,(\mu,\rm etc.=1,2,3,4)$ where $H^{\nu\rho\sigma}$ is the field strength of $B^{\rho\sigma}$  \cite{Witten84}. In this compactification with U(1)$_A$,   $ {A}_\mu$  absorbs  $a_{\rm MI}$ to become massive at the  compactification scale $m_A\approx 10^{18\,}\gev$.\footnote{If there is no anomalous U(1) gauge symmetry, the MI axion is an axion at $m_{\rm MI}\approx 10^{16\,}\gev$ \cite{ChoiKim85}.} Below the scale $m_A$, there remains a global symmetry which is called \Uanom, and its charge is given by $\Qanom$ presented in Eq. (\ref{eq:Qanom}).
 In detail, it works as follows.  Suppose that five $\tilde{\rm U}(1)$ charges out of $Q_{1,\cdots,6}$ are broken, and there is only one gauge symmetry remaining, which we identify as \Uanom.   Now, we can consider two continuous parameters, one is the MI-axion direction and the other the phase of \Uanom~transformation. Out of two continuous directions, only one phase or pseudoscalar is absorbed by the U(1)$_{anom}$ gauge boson, and one continuous direction survives.
The remaining continuous degree corresponds to a global symmetry, which  is called the 't Hooft mechanism \cite{Hooft71}: ``If both a gauge symmetry and a global symmetry are broken by one scalar vacuum expectation value (VEV), the gauge symmetry is broken and a global symmetry is surviving''.  The resulting global charge is a linear combination of the original gauge and global charges. 
We will briefly review this in Appendix.
 This counting of pseudoscalar degrees is not affected by changing the scales of the VEVs. So, when the anomalous U(1) is arising at the compactification scale, the gauge symmetry  \Uanom~is broken and the MI-axion degree is removed, and in addition there results the global  \Uanom~symmetry below the compactification scale. The dilaton the partner of the MI-axion must remain heavy   as in the usual Higgs mechanism because it does not find its partner below the compactification scale. 
 
 \subsection{No gauged anomalous U(1) below the compactification scale}
 
\begin{figure}[!t]
\begin{center}
\includegraphics[width=0.6\textwidth]{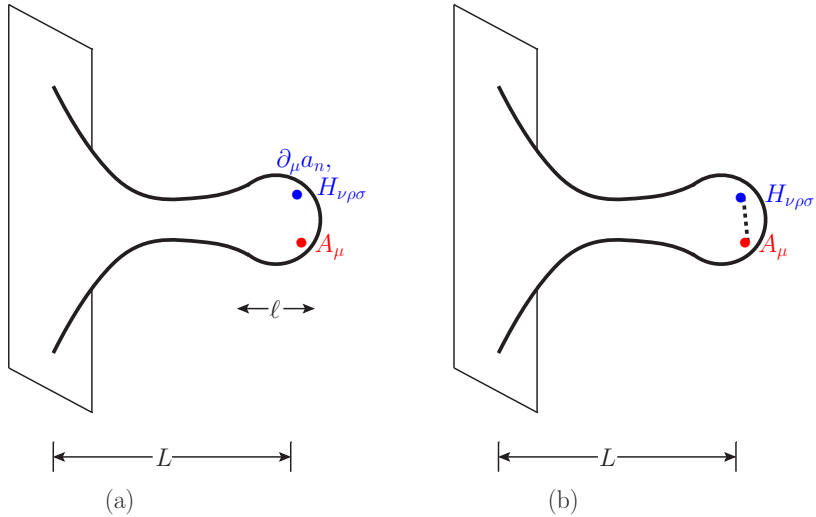} 
\end{center}
\caption{Compactification, leading to $M_4\times K$. The parallelogram depicts $M_4$, $\ell$ is a compactification size of $K$, and $L$ in one direction is shown pictorially as a neck.} \label{fig:M4K}
\end{figure}

There have been many discussions on the Fayet-Iliopoulos (FI) $D$-term for \Uanom~gauge symmetry at the GUT scale, e.g. in Ref. \cite{DineSW87}. However, there is no gauged U(1) corresponding to the anomalous symmetry below the compactification scale. In models with a hierarchy between the compactification scale and the GUT scale, so there is no need to consider the  \Uanom~ $D$-term, $\int d^4\theta \,\xi D$, below the compactification scale. The $\xi$ term is in the D term potential, $\frac12 D^2$, with $D=-\xi-e\phi^*Q_a\phi$ where $e$ is the \Uanom~ gauge coupling.   Our \Uanom~is derived from the orbifold compactification of the $\EE8$ heterotic string.  After compactification of the six internal space, one can consider $M_4\times K$ where $M_4$ is the Minkowski space and $K$ is the internal space. In Fig. \ref{fig:M4K}, we show some relevant fields living in $K$. The effective symmetry group in the $M_4$ Lagrangian is gauge symmetries times some discrete groups without any global symmetry except that corresponding to $B_{\mu\nu}$. The 4D scalar $B_{ij}$ are called the  MD-axions, $a_n\
(n\ge 1)$, determined by the topology of the internal space. Reference \cite{DineSW87} shows that the classical symmetries corresponding to $a_n\to a_n+({\rm constant})$ are broken by the world-sheet instanton effects.  Fig. \ref{fig:M4K}\,(a) shows these fields living in $K$ with some U(1) gauge fields. In the compactification of $\EE8$ heterotic string, there appears only one anomalous U(1) if any such terms are present. If so, the corresponding gauge boson obtains mass by absorbing $a_{\rm MI}$ as shown in Appendix and  Fig. \ref{fig:M4K}\,(b). All the other U(1)'s are anomaly free. $a_n$ are not absorbed to gauge bosons at this stage. So, we can consider the massless states $a_n$ and non-anomalous six U(1)'s below the compactification scale, and their K$\ddot{\rm a}$hler potentials and FI  D terms. However, for \Uanom~ we need not consider the corresponding D term.  

For a consistency check, consider  Fig. \ref{fig:M4K}\,(a) again. If the throat is not cut, the space is still 10D. In this 10D, we can consider the \Uanom~subgroup of $\EE8$. Let us consider the subgroup \smfa$\times \prod_{i=1}^5\tilde{\rm U}(1)_i$ of $\EE8$ where $\tilde{\rm U}(1)_i$ are anomaly free. Before cutting the throat, let us break $\EE8$ to separate out \Uanom~by the VEV of an appropriate adjoint representation (in 4D language) in the bulk. Of course, this adjoint representation is not present in our massless spectra but good enough to see the resulting effective low energy theory. If \Uanom~is separated out in this way, it obtains mass by absorbing  the MI axion by the VEVs of $F_{ij}$ as shown in Appendix. Then, assign superheavy masses to the adjoint representation we introduced. So far, nothing has been introduced violating the effective symmetry \smfa$\times \prod_{i=1}^5\tilde{\rm U}(1)_i$. Next, cut the throat to obtain the effective 4D theory which is \smg$\times\prod_{i=1}^5\tilde{\rm U}(1)_i$. There cannot be a FI D term for \Uanom~in this interpretation.
 
Even if we consider the FI term with a non-vanishing $\xi$, in case there is no hierarchy between the compactification scale and the GUT scale,\cite{AtickTwoLoop88} we can show that a global symmetry can be derived below the scale of the anomalous gauge boson mass. For this, let us consider Eq. (\ref{eq:HooftMI}) plus the D term with a non-vanishing $\xi$,
\dis{
\frac12 \partial^\mu  a_{\rm MI}\partial_\mu a_{\rm MI} &+    M_{\rm MI}A_\mu\partial^\mu a_{\rm MI} +\left|- \xi+e\sum_{a}\phi_a^*Q_a \phi_a\right|^2+\Big[|(\partial_\mu -ie A_\mu)\phi_1|^2+ \cdots\Big]\\
&=  (M_{\rm MI} \partial^\mu a_{\rm MI}-eV_1\partial^\mu a_1)A_\mu+\cdots,  \label{eq:TwoAtGUT}
}
where $\phi_a$ are assumed to carry only the anomalous charge for a moment, not carrying any non-anomalous charge $Y$ and $\tilde{Q}_i\,(i=1,\cdots,5)$.   Let one $\phi_a$, say $\phi_1$ develops a VEV, ${V_1}$, by minimizing the FI term. Here, two phase fields, $a_{\rm MI}$ and $a_1$ [= the phase of $\phi_1\,(=(V_1+\rho_1)e^{ia_1/V_1})/\sqrt2$] are considered and only one Goldstone boson is absorbed to $A_\mu$,
\dis{
\sqrt{M_{\rm MI}^2+e^2V_1^2}\, (\cos\theta_G\, a_{\rm MI}-\sin\theta_G\, a_1)\label{eq:LongA}
 } 
where
$\tan\theta_G=eV_1/M_{\rm MI}$. The orthogonal direction
\dis{
\theta'=\cos\theta_G\, \frac{a_{\rm MI}}{|M_{\rm MI}|}+\sin\theta_G\,\frac{a_1}{|eV_1|}\label{eq:globalD}
}
is surviving as a global direction below the scale $\sqrt{M_{\rm MI}^2+e^2V_1^2}$.
With this global symmetry, we can consider breaking the five non-anomalous $\tilde{\rm U}(1)$'s around the GUT scale, leaving only one global symmetry to the axion window.
If one gauge symmetry  \Uanom~were the whole story for the global symmetry, the next lower scale VEV of a scalar carrying non-zero VEV (probably at a GUT scale) will break the global \Uanom.
In the orbifold compactification, however, there appear many gauge U(1)'s (six in our example) which are anomaly free except the above \Uanom, say U(1)$_4$ in our example. Now, the D term of   U(1)$_4$ is $|\phi_2^*Q_4\phi_2|^2$. Generally, $\phi_2$ carries the \Uanom~charge, for example in Table \ref{tb:singlets} any $\phi_2$ carrying non-zero $Q_4$ also carries $\Qanom$. So, the VEV of $\phi_2$ will break both  U(1)$_4$ and the global \Uanom~obtained above. Below $\langle \phi_2\rangle$, there appears the global symmetry \Uanom. So, applying the 't Hooft mechanism repeatedly until all anomaly free gauge U(1)'s are broken except U(1)$_Y$, we obtain the global \Uanom~interpretable as \UPQ~from string compactification.\footnote{It is easy to see this by counting the number of continuous degrees. We introduced two phases, $a_{\rm MI}$ and $a_1$. One combination is absorbed to  $A^\mu_{\rm anom}$. One may worry that the other combination, (\ref{eq:globalD}), might become  massive because we considered two terms for  $a_{\rm MI}$ and $a_1$. But, it does not work that way because both generators, \Uanom~and  $Q_a$ in the FI D-term in Eq. (\ref{eq:TwoAtGUT}), are proportional.   Eq. (\ref{eq:LongA}) explicitly shows that only one combination  is removed to   $A^\mu_{\rm anom}$.}
 Then, at the intermediate scale $10^{9-11}$ GeV, one \Uanom~ breaking VEV $f_a$ of a SM singlet scalar $\Phi$ breaks the global symmetry \Uanom~spontaneously and there results the needed ``invisible''
axion at the intermediate scale. 
The global symmetry whose shift angle $\theta'$ is broken at an intermediate scale to create the
 ``invisible'' axion by some scalar field carrying the anomalous charge. To observe this, only one scalar field  $\phi_1$ develops a VEV $V_1$ as before. Now let us define a global charge surviving below the scale $\sqrt{M_{\rm MI}^2+e^2V_1^2}$ as\footnote{The sign in front of $\Qanom$ belongs to the sign convention of the $\overline{\theta}$ term and we choose + sign here.}
\dis{
\hat{Q}'={Q}_{\rm anom}+x\,{Q}_a\label{eq:NewQanom}
}
where $ {Q}_a$ is an anomaly free gauge
U(1) charge. The condition $\hat{Q}'|\phi_1\rangle=0$ so that $\hat{Q}'$ is a good generator of the global symmetry, we fix $x=-{Q}_{\rm anom}(\phi_1)/Q_a(\phi_1)$ . Since $Q_a$ is an anomaly-free generator, we can add a constant multiple of $Q_a$ in Eq. (\ref{eq:NewQanom}) to give the same anomaly coefficient for U(1)$'$--SU(3)$_c$--SU(3)$_c$. So, we can take a new global charge as
\dis{
\hat{Q}'={Q}_{\rm anom}.\label{eq:NewQ}
}
Actually, a more general proof was given in Ref. \cite{KimPRD17}. Consider $\hat{Q}'={Q}_{\rm anom}+x_1 Q_a+x_2Q_b$ where $Q_a$ and $Q_b$ are anomaly-free gauge charges. Then, Tr\,$\hat{Q}'=\Qanom$ and Tr\,$\hat{Q}'Q_pQ_q=\Qanom Q_pQ_q$ where $\{p,q\}=\{a,b\}$. In fact, in Ref.  \cite{KimPRD17}, instead of $\Qanom$ of Eq. (\ref{eq:Qanom}), we have shown exactly the same traces  with
\dis{
\hat{Q}'=63(0^5;16,28,-8)(0^5;6,6,-12)' .\label{eq:Qanompr}
}
 In the Tables of the present paper   $\Qanom$ is used, and in the Tables of \cite{KimPRD17} $\hat{Q}'$ is used, but the anomaly-related quantities such as Tr\,$\hat{Q}'Q_{\rm color}^\alpha Q^\beta_{\rm color}$ and  Tr\,$\hat{Q}'\Qem \Qem$ are exactly the same in both cases.

\subsection{``Invisible'' axion in the axion window}

For simplicity, assume a hierarchy between the compactification and the GUT scales. Let us choose $\phi$, not carrying any gauge charge. The VEV of $\phi$ is assumed to be at the axion window and breaks the \Uanom~global symmetry.
In this case, the actual global symmetry breaking scale  is a mixture of two effects: the MI-axion direction and the hypothetical intermediate scale axion direction (the phase of $\phi$).      Let the original anomalous gauge charge of $\phi$ be $Q_a$, $\phi\,(\sim ve^{ia_\phi/v})$. Then, the QCD axion $a$ created at the intermediate scale (determined by the VEV of $\phi$) is a combination of $a_\phi$ and $a_{\rm MI}$,
\dis{
a=\cos\theta \,a_\phi+\sin\theta\,a_{\rm MI},~~{\rm with}~\sin\theta=
\frac{gQ_av}{\sqrt{M^2_{\rm MI}+g^2Q_a^2v^2}}
}
where the antisymmetric tensor field strength is the MI-axion, $H_{\mu\nu\rho}=M_{MI}\epsilon_{\mu\nu\rho\sigma} \partial^\sigma a_{MI}$ \cite{Witten84}. Thus, for $v\ll M_{MI}$ we obtain the desired ``invisible'' axion at the intermediate scale.  

Here, we stress again that the exact global symmetries from string compactification require anomalous gauge symmetries after compactification. From the $\EE8$ heterotic string there is only one such anomalous gauge symmetry we discussed above.\footnote{There can be more anomalous gauge symmetries from Type-I and Type II-B \cite{Uranga99}.} Any other global symmetries must be accidental as discussed for QCD axions in \cite{ChoiAcc,Accions} and for axion-like particles in \cite{Vaudrevange14}.

\subsection{Three families in the flipped SU(5)}

We require that there result the three families of the minimal supersymmetric standard model (MSSM) and one pair of Higgs doublets. To have the ``invisible'' axion, we further require that \Uanom~is broken at the intermediate scale, $\Mi\approx 10^{11\,}\gev$. At the   SU(5)$_{\rm flip}$  GUT level, we need three copies of $\overline{\ten}_{-1}\oplus {\five}_{+3}\oplus{\one}_{-5}$. In Table \ref{tb:colorfields}, we present SU(5) non-singlet fields with the global quantum numbers where the axion-photon-photon couplings are presented in the last column. One family appears in the untwisted sector $U$, and two families appear in the twisted sector $T_4^0$.
 
In the SU(5)$_{\rm flip}$ GUT, one needs a pair  $\tenb_{-1}\oplus{\ten}_{+1}$ for  breaking the rank 5 group SU(5)$_{\rm flip}$ down to the rank 4 group 
\smg. Indeed, they appear in $T_3^0$ and $T_9^0$ in Table \ref{tb:colorfields}. Also, the vacuum expectation values (VEVs) of these pairs achieve the doublet-triplet splitting discussed in Subsec. \ref{subsec:DT}.

\begin{table}[!t]
{\tiny
\begin{center}
\begin{tabular}{|c|c|c |c||ccccccc|c|c|}
\hline Sect. & Colored states & SU(5)$_X$ &
Mult. & $Q_1$& $Q_2$ & $Q_3$ & $Q_4$ & $Q_5$ & $Q_6$ &  $Q_{\rm anom}$ & Label &$Q_a^{\gamma\gamma}$ \\[0.3em]
\hline\hline

$U$ & $\left(\underline{+\,+\,+\,-\,-\,};-\,-\,+\,
\right)(0^8)'$ & $\tenb_{-1}$     &  &--6 &--6 &+6 &0 &0 &0 &$-1638$\textbf{(-13)} & $ C_2$&$-3276$ \\[0.3em]
$U$ & $\left(\underline{+\,-\,-\,-\,-\,};+\,-\,-\,
\right)(0^8)'$ & $\five_{+3}$  &    &+6 &--6 &--6 &0 &0 &0 &$-126$\textbf{(-1)}  & $C_1$&$-294$ \\[0.3em]
\hline

$T_{4}^0$ &
$\left(\underline{+\,-\,-\,-\,-\,};\frac{-1}{6}\,\frac{-1}{6}\,\frac{-1}{6}\,
\right)(0^8)'$ &  $\five_{+3}$   & $2$ &$-2$ &$-2$ &$-2$ &0 &0 &0 &$-378$\textbf{(-3)} & $2C_3$&$-882$ \\[0.5em]
$T_{4}^0$ &
$\left(\underline{+\,+\,+\,-\,-\,};\frac{-1}{6}\,\frac{-1}{6}\,\frac{-1}{6}\,
\right)(0^8)'$ &   $\tenb_{-1}$ & $2$ &$-2$ &$-2$ &$-2$ &0 &0 &0 &$-378 $\textbf{(-3)} & $ 2C_4$ &$-756$\\[0.5em]
\hline

$T_{4}^0$ &
$\left(\underline{1\,0\,0\,0\,0\,};\frac{1}{3}\,\frac{1}{3}\,\frac{1}{3}\,
\right)(0^8)'$ &  $\five_{-2}$   & $2$ &$+4$ &$+4$ &$+4$ &0 &0 &0 &$+756$\textbf{(+6)}  & $2C_5$&$+1008$ \\[0.5em]
$T_{4}^0$ &
$\left(\underline{-1\,0\,0\,0\,0\,};\frac{1}{3}\,\frac{1}{3}\,\frac{1}{3}\,
\right)(0^8)'$ &   $\fiveb_{+2}$ & $2$ &$+4$ &$+4$ &$+4$ &0 &0 &0 &$+756$\textbf{(+6)}  & $2C_6$ &$+1008$ \\[0.5em]
\hline

$T_{6}^0$ &
$\left(\underline{1\,0\,0\,0\,0\,};0\,0\,0\,
\right)(0^5; \frac{-1}{2}\,\frac{+1}{2}\,0)'$ &  $\five_{-2}$   & $3$ &0 &0 &0 &$-12$ &0 &0 &0
 & $3C_7$&0 \\[0.5em]
$T_{6}^0$ &
$\left(\underline{-1\,0\,0\,0\,0\,};0\,0\,0\,
\right)(0^5; \frac{+1}{2}\,\frac{-1}{2}\,0)'$ &   $\fiveb_{+2}$  & $3$ &0 &0 &0 & $+12$ &0 &0 &
0 & $3C_8$&0 \\[0.5em]
\hline

$T_{7}^0$ &
$\left(\underline{-1\,0\,0\,0\,0\,};\frac{-1}{6}\,\frac{-1}{6}\,\frac{-1}{6}\,
\right)(0^5; \frac{-1}{4}\,\frac{-1}{4}\,\frac{+2}{4})'$ &  $\fiveb_{+2}$ & $1$ &$-2$ &$-2$ &$-2$ &0&$+9$ & $+3$&$-972$\textbf{(-$\frac{54}{7}$)} & $C_{9}$ &$-1296$ \\[0.5em]
$T_{7}^0$ &
$\left(\underline{+1\,0\,0\,0\,0\,};\frac{-1}{6}\,\frac{-1}{6}\,\frac{-1}{6}\,
\right)(0^5; \frac{-1}{4}\,\frac{-1}{4}\,\frac{+2}{4})'$ &   $\five_{-2}$ & $1$ &$-2$ &$-2$  &$-2$ &0 &$+9$ &$+3$ &$-972$\textbf{(-$\frac{54}{7}$)}& $C_{10}$&$-1296$ \\[0.5em]
\hline

$T_{3}^0$ &$\left(\underline{+\,+\,+\,-\,-\,};0\,0\,0\,
\right)(0^5; \frac{-1}{4}\,\frac{-1}{4}\,\frac{+2}{4})'$ &  $\tenb_{-1}$ & $1$ &0 &0 &0 &0 &$+9$ &$+3$ &$-594$\textbf{(-$\frac{33}{7}$)}& $ C_{11}$&$-1188$  \\[0.5em]
$T_{9}^0$ &
$\left(\underline{+\,+\,-\,-\,-\,};0\,0\,0\,
\right)(0^5; \frac{+1}{4}\,\frac{+1}{4}\,\frac{-2}{4} )'$ &   $\ten_{+1}$ & $1$ &0 &0 &0 &0 &$-9$ &$-3$ &$+594$\textbf{(+$\frac{33}{7}$)} & $ C_{12}$&$+1188$ \\[0.5em]
\hline\hline
& &  & &$-16$ &$-28$  &$+8$ &0 &$+18$ &$+6$&$-3492\qquad$ &  &$ -5406$  \\[0.3em]
\hline

\end{tabular}
\end{center}
\caption{The \SUflip~ states. Here, + represents helicity $+\frac12$ and -- represents helicity $-\frac12$. Sum of $Q_{\rm anom}$ is multiplied by the index of the fundamental representation of SU(3)$_c$, $\frac12$.  The PQ symmetry, being chiral, counts quark and antiquark in the same way.  The right-handed states in $T_3$ and $T_5$ are converted to the left-handed ones of $T_9$ and $T_7$,
respectively. The bold entries are $\Qanom/126$.
}\label{tb:colorfields} }
\end{table}
 
For Higgs quintets, $T_6$ has the pairs with multiplicity 3.
More importantly,  two pairs  appear in  $T_4^0$, and one pair appears in $T_7^0$.
 Two pairs  appearing in  $T_4^0$ are not distinguished, and  the Higgsino mass matrix elements are democratic.
     Their Yukawa couplings take the form $C_5C_{6}\sigma_{1}$, which conserves the \Uanom~symmetry,
\begin{eqnarray}
M_{\rm demo}^{2\times 2}= \begin{pmatrix}
M&M \\ M&M 
\end{pmatrix}\label{eq:Mdemo2}
\end{eqnarray}
where $M\sim \langle\sigma_{1}\rangle$. The $\sigma_1$ multiplicity  is 3, as shown   in Table \ref{tb:singlets}. The twisted sector $T_4^0$ satisfies $\Z_3$ orbifold selection rules and the multiplicity 3 belonging to the permutation symmetry $S_3$  splits into  $S_3$ representations $\two\oplus\one$. Three $\sigma_1$'s under $S_3$ in the $\Z_3$ compactification can be combined to \cite{ChoiKS03,Segre79}
\dis{
&\Phi_0=  \frac{1}{\sqrt3} \left(\sigma_1^{a}+
\sigma_1^{b}+\sigma_1^{c} \right) ,\\
&\Phi_+=\frac{1}{\sqrt3}\left(\sigma_1^{a}+
\omega\,\sigma_1^{b}+ \bar\omega\,\sigma_1^{c} \right) ,\\
&\Phi_-=\frac{1}{\sqrt3}\left(\sigma_1^{a}+{\bar\omega}\,\sigma_1^{b}
+\omega\,\sigma_1^{c} \right),
}
where $\omega=e^{2\pi i/3}$ and ${\bar\omega}=e^{4\pi i/3}$ are the cube roots of unity. $\Phi_0$ is a singlet {\bf 1} and $\Phi_+$ and $\Phi_-$ form a doublet {\bf 2}. Suppose that $\langle\Phi_0\rangle\ne 0$ and $\langle\Phi_+\rangle=\langle\Phi_-\rangle= 0$. $C_5$ and $C_6$ belong  to {\bf 2} of $S_3$, and their multiplication gives ${\two}\times{\two}={\one}\oplus{\one}'
\oplus{\two}$,
\dis{
&\Psi_0=  \frac{1}{\sqrt3} \left(C_5^{(1)}C_6^{(1)}
+C_5^{(2)}C_6^{(2)}\right) ,\\
&\Psi_0'=  \frac{1}{\sqrt3} \left(C_5^{(1)}C_6^{(1)}
-C_5^{(2)}C_6^{(2)}\right) ,\\
&\Psi_+=  \frac{1}{\sqrt3} \left(C_5^{(1)}C_6^{(2)}
+\omega\,C_5^{(2)}C_6^{(1)}\right) ,\\
&\Psi_-=  \frac{1}{\sqrt3} \left(C_5^{(1)}C_6^{(2)}
+\bar\omega\,C_5^{(2)}C_6^{(1)}\right) ,\\
 }
 where $\Psi_+$ and $\Psi_-$ form a doublet under interchange $(1)\leftrightarrow (2)$. Thus, the singlet VEV $\langle\Phi_0\rangle$ can couple either $\Psi_0$ or $\Psi_0'$. In this way, one pair becomes superheavy.
 This result is equivalent to the democratic mass matrix (\ref{eq:Mdemo2}).
Namely, determinant of $M_{\rm demo}^{2\times 2}$ is 0, and only one pair obtains mass $2M$. The remaining pair is massless at this stage.
For the three pairs in $T_6$, the Higgsino mass matrix can  be studied in the same way. Since it belongs to $T_6$, we consider $\Z_2$ and $S_2$ permutation which allow only the following singlet combinations,
\dis{
&\Psi^{(a)}=  \frac{1}{\sqrt3} \left(C_7^{(1)}C_8^{(1)}
+C_7^{(2)}C_8^{(2)}+C_7^{(3)}C_8^{(3)}\right) ,\\
&\Psi^{(b)}=  \frac{1}{\sqrt3} \left(C_7^{(1)}C_8^{(1)}
-C_7^{(2)}C_8^{(2)}+C_7^{(3)}C_8^{(3)}\right) ,\\
&\Psi^{(c)}=  \frac{1}{\sqrt3} \left(C_7^{(1)}C_8^{(1)}
+C_7^{(2)}C_8^{(2)}-C_7^{(3)}C_8^{(3)}\right) ,\\
&\Psi^{(d)}=  \frac{1}{\sqrt3} \left(C_7^{(1)}C_8^{(1)}
-C_7^{(2)}C_8^{(2)}-C_7^{(3)}C_8^{(3)}\right).
 }
 If we require the invariance of the mass matrix under $S_2$, \ie under the interchange of any two pairs out of $(1), (2)$ and $(3)$, only the term $\Psi^{(a)}$ is allowed. Then, only  one pair obtains superheavy mass and two pairs remain light. Again it is like taking a democratic mass matrix.
   Even if two pairs from $T_6$ remain light, their contribution to the unification point of couplings is null because they are  the SU(5)$_{\rm flip}$ complete multiplets.  However, the absolute magnitude of the gauge coupling constant at the unification point is affected. Nevertheless we will not discuss these complete multiplets any more in this paper  since the massless pairs do not affect our discussion on the flavor problems.

There are Yukawa couplings
$C_5C_{11}\sigma_{21}$ and $C_6C_{12}\sigma_{21}$, which conserves the \Uanom~symmetry. So, among the remaining two light pairs (one from $T_4^0$ and the other from $T_7^0$), one obtains mass, and finally there will be left with only one light pair. The $3\times 3$ Higgsino mass matrix  is,\footnote{In the (33) position, $\varepsilon$ will be introduced  later.}
\begin{eqnarray}
&& ~~H_u^{(T4)_1}~~\,  H_u^{(T4)_2}~~ H_u^{(T7)}\nonumber  \\
 M_{\rm Higgsino}^{3\times 3}&=&\begin{pmatrix}
 ~~M~&~~\quad M\quad &\quad m~~\\[0.5em]  
 ~~M~&~~\quad M\quad &\quad m~~\\[0.5em]  ~~m~&~~\quad m\quad &\quad 0~~
\end{pmatrix} \begin{array}{c}H_d^{(T4)_1}\\[0.5em] H_d^{(T4)_2}\\[0.5em]  H_d^{(T7)} \end{array}\label{eq:Mhigginos}
\end{eqnarray}
where $m\sim \langle\sigma_{21}\rangle$. As expected, determinant of $M_{\rm Higgsino}^{3\times 3}$ is 0, and there remain two  light pairs as far as the (33) element is 0. The heaviest eigenstate of (\ref{eq:Mhigginos}) is
\begin{eqnarray}
\Psi^{M_c}=\frac{1}{\sqrt2}\left(\Psi^{T_4^0}_1+\Psi^{T_4^0}_2\right), ~{\rm mass}=2M,
\end{eqnarray}
where $\Psi$ is $H_{u,d}$. The Higgsino pair of the MSSM contains
\begin{eqnarray}
\Psi^{0}=\frac{1}{\sqrt2}\left(\Psi^{T_4^0}_1-\Psi^{T_4^0}_2\right), ~{\rm mass}=0.\label{eq:Massless}
\end{eqnarray}
The other  state  with a nonzero (33) element will be presented later.
 
\begin{table}[!t]
{\tiny
\begin{center}
\begin{tabular}{|c|c|c|c|c||ccccccc|c|}
\hline sectors & Neutral singlet states &   SU(5)$_X$  & $(N^L)_j$ &
${\cal P}(f_0)$&$Q_1$ &$Q_2$ &$Q_3$ &$Q_4$ &$Q_5$ &$Q_6$ &$Q_{\rm anom}$   & La. \\[0.2em]
\hline\hline

$T_{4}^0$ &
$\left(0^5~;\frac{-2}{3}~\frac{-2}{3}~\frac{-2}{3}
\right)(0^8)'$ &$\one_{0}$ & $0$ & $3$   &$-8$ &$-8$ &$-8$ &0&0&0&$-1512$\textbf{(-12)} & $\sigma_1$ \\[0.3em]
$T_{4}^0$ & $\left(0^5~;\frac{-2}{3}~\frac{1}{3}~\frac{1}{3}
\right)(0^8)'$ &$\one_{0}$ & $1_{\bar 1},1_{2},1_{3}$ & $2,3,2$ &$-8$ &$+4$ &$+4$ &0&0&0&$-252$\textbf{(-2)}  & $\sigma_2$ \\
$T_{4}^0$ & $\left(0^5~;\frac{1}{3}~\frac{-2}{3}~\frac{1}{3}
\right)(0^8)'$ &$\one_{0}$ & $1_{\bar 1},1_{2},1_{3}$ & $2,3,2$
&$+4$ &$-8$ &$+4$ &0&0&0& $-1008$\textbf{(-8)}   & $\sigma_3$ \\[0.3em]
$T_{4}^0$ & $\left(0^5~;\frac{1}{3}~\frac{1}{3}~\frac{-2}{3}
\right)(0^8)'$ &$\one_{0}$ & $1_{\bar 1},1_{2},1_{3}$ & $2,3,2$ &$+4$ &$+4$ &$-8$ &0&0&0&$+1260$\textbf{(+10)}  & $\sigma_4$ \\[0.3em]
\hline

$T_{6}$ &  $\left(0^5~;0~ 1~0\right)
 \left(0^5~\frac{1}{2}~\frac{-1}{2}~0 \right)'$ &
$\one_{0}$ & $0$ &  $2$
&0&$+12$ &0&$+12$ &0&0&$+1764$\textbf{(+14)}   & $\sigma_5$ \\[0.3em]
$T_{6}$ &  $\left(0^5~;0~0 ~1\right)
 \left(0^5~\frac{-1}{2}~\frac{1}{2}~0 \right)'$ &
$\one_{0}$ & $0$ &  $2$  &0&0&$+12$ &$-12$ &0&0&$-504$\textbf{(-4)}  & $\sigma_6$\\
$T_{6}$ &  $\left(0^5~;0~ -1~0\right)
 \left(0^5~\frac{-1}{2}~\frac{1}{2}~0 \right)'$ &
$\one_{0}$ & $0$ &  $2$
 &0&$-12$ &0&$-12$ &0&0&$-1764$\textbf{(-14)}   & $\sigma_7$ \\[0.3em]
$T_{6}$ &  $\left(0^5~;0~0 -1\right)
 \left(0^5~\frac{1}{2}~\frac{-1}{2}~0 \right)'$ &
$\one_{0}$ & $0$ & $2$ &0&0&$-12$ &$+12$ &0&0&$+504$\textbf{(+4)} & $\sigma_8$
\\[0.3em]
\hline

  $T_{2}^0$ &
$\left(0^5~;\frac{-1}{3}~\frac{-1}{3}~\frac{-1}{3}
\right)(0^5~\frac{-1}{2}~\frac{1}{2}~0)'$ &$\one_{0}$ & $2_{\bar
1},2_{3}$ & $1,1$ &$-4$ &$-4$ &$-4$ &$-12$ &0&0&$-756$\textbf{(-$6$)} & $\sigma_{9}$ \\[0.3em] 
$T_{2}^0$ &
$\left(0^5~;\frac{-1}{3}~\frac{-1}{3}~\frac{-1}{3}
\right)(0^5~\frac{1}{2}~\frac{-1}{2}~0)'$ &$\one_{0}$ & $2_{\bar
1},2_{3}$ & $1,1$
&$-4$ &$-4$ &$-4$ &$+12$ &0&0&$-756$\textbf{(-$6$)}  & $\sigma_{10}$ \\[0.3em] \hline

$T_{3}$ &
$\left(0^5~;\frac{-1}{2}~\frac{-1}{2}~\frac{-1}{2}
 \right)\left(0^5~\frac{3}{4} ~\frac{-1}{4}~ \frac{-1}{2}\right)'$ &$\one_{0}$ 
 & $0$  & $1$ &$-6$ &$-6$ &$-6$ &$+12$&$-9$ &$-3$&$-540$\textbf{(-$\frac{30}{7}$)} & $\sigma_{11}$ \\[0.3em]
$T_{3}$ &  $\left(0^5~;\frac{-1}{2}~\frac{1}{2}~\frac{1}{2}
 \right)\left(0^5~\frac{3}{4} ~\frac{-1}{4}~ \frac{-1}{2}\right)'$ &$\one_{0}$ 
 & $0$  & $1$ &$-6$ &$+6$ &$+6$ &$+12$&$-9$ &$-3$&$+720$\textbf{(+$\frac{40}{7}$)} & $\sigma_{12}$ \\[0.3em]
$T_{3}$ &  $\left(0^5~;\frac{1}{2}~\frac{1}{2}~\frac{-1}{2}
 \right)\left(0^5~\frac{-1}{4} ~\frac{3}{4}~ \frac{-1}{2}\right)'$ &$\one_{0}$ 
 & $0$  & $1$ &$+6$ &$+6$ &$-6$ &$-12$&$-9$ &$-3$&$+2232$\textbf{(+$\frac{124}{7}$)} & $\sigma_{13}$ \\[0.3em]
$T_{3}$ &  $\left(0^5~;\frac{1}{2}~\frac{1}{2}~\frac{-1}{2}
 \right)\left(0^5~\frac{-1}{4} ~\frac{-1}{4}~ \frac{1}{2}\right)'$ &$\one_{0}$ 
 & $1_{1},1_{3}$ &  $2$,$1$
&$+6$ &$+6$ &$-6$ &0&$+9$ &$+3$&$+1044$\textbf{(+$\frac{58}{7}$)}  & $\sigma_{14}$\\[0.3em]
\hline $T_{9}$ &
$\left(0^5~;\frac{1}{2}~\frac{1}{2}~\frac{1}{2}
 \right)\left(0^5~\frac{-3}{4} ~\frac{1}{4}~ \frac{1}{2}\right)'$ &$\one_{0}$ 
 & $0$  & $1$ &$+6$ &$+6$ &$+6$ &$-12$&$+9$ &$+3$&$+540$\textbf{(+$\frac{30}{7}$)} & $\sigma_{15}$ \\[0.3em]
$T_{9}$ &  $\left(0^5~;\frac{1}{2}~\frac{-1}{2}~\frac{-1}{2}
 \right)\left(0^5~\frac{-3}{4} ~\frac{1}{4}~ \frac{1}{2}\right)'$ &$\one_{0}$ 
 & $0$ & $2$
&$+6$ &$-6$ &$-6$ &$-12$&$+9$ &$+3$& $-720$\textbf{(-$\frac{40}{7}$)} & $\sigma_{16}$ \\[0.3em]
$T_{9}$ &  $\left(0^5~;\frac{-1}{2}~\frac{-1}{2}~\frac{1}{2}
 \right)\left(0^5~\frac{1}{4} ~\frac{-3}{4}~ \frac{1}{2}\right)'$ &$\one_{0}$ 
 & $0$  & $2$ &$-6$ &$-6$ &$+6$ &$+12$&$+9$ &$+3$&$-2232$\textbf{(-$\frac{124}{7}$)} & $\sigma_{17}$ \\[0.3em]
$T_{9}$ &  $\left(0^5~;\frac{-1}{2}~\frac{-1}{2}~\frac{1}{2}
 \right)\left(0^5~\frac{1}{4} ~\frac{1}{4}~ \frac{-1}{2}\right)'$ &$\one_{0}$ 
 & $1_{\bar 1}$,$1_{\bar 3}$ &  $1$,$1$
&$-6$ &$-6$ &$+6$ &0&$-9$ &$-3$& $-1044$\textbf{(-$\frac{58}{7}$)} & $\sigma_{18}$ \\[0.3em]
\hline

$T_{1}^0$ &
$\left(0^5~;\frac{-1}{6}~\frac{-1}{6}~\frac{-1}{6}
\right)(0^5~\frac{-3}{4}~\frac{1}{4}~\frac12)'$ &$\one_{0}$  & $3_{3}$ & $1$
 &$-2$ &$-2$ &$-2$ &$-12$ &$+9$ &$+3$& $-972$\textbf{(-$\frac{54}{7}$)}& $\sigma_{19}$ \\[0.3em]
$T_{1}^0$ &
$\left(0^5~;\frac{-1}{6}~\frac{-1}{6}~\frac{-1}{6}
\right)(0^5~\frac{1}{4}~\frac{-3}{4}~\frac12)'$ &$\one_{0}$ &$3_{3}$ & $1$
&$-2$ &$-2$ &$-2$ &$+12$ &$+9$ &$+3$&  $-972$\textbf{(-$\frac{54}{7}$)}& $\sigma_{20}$ \\[0.3em]
$T_{1}^0$ &
$\left(0^5~;\frac{-1}{6}~\frac{-1}{6}~\frac{-1}{6}
\right)(0^5~\frac{1}{4}~\frac{1}{4}~\frac{-1}{2})'$ &$\one_{0}$ &
 $\{1_{1},1_{3}\}$ & $1$ &$-2$ &$-2$ &$-2$ &0&$-9$ &$-3$&  $+216$\textbf{(+$\frac{12}{7}$)}& $\sigma_{21}$ \\[0.3em]
 & & &$\{2_{3},1_{2}\}$  & $1$ &  &  & && & &   &   \\[0.3em]
&  & &$6_{3}$ & $1 $ &  & & & & & & &   \\[0.3em]
\hline

$T_{7}^0$ &  $\left(0^5~;
\frac{5}{6}~\frac{-1}{6}~\frac{-1}{6}\right)
 (0^5~\frac{-1}{4}~\frac{-1}{4}~\frac{1}{2})'$ & $\one_{0}$  & $2_{\bar 1}$ &$1$
&$+10$ &$-2$ &$-2$ &0&$+9$ &$+3$&$+36$\textbf{(+$\frac{2}{7}$)}  & $\sigma_{22}$  \\[0.3em]
$T_{7}^0$ &  $\left(0^5~;
\frac{-1}{6}~\frac{5}{6}~\frac{-1}{6}\right)
 (0^5~\frac{-1}{4}~\frac{-1}{4}~\frac{1}{2})'$ & $\one_{0}$ & $2_{\bar 1}$ &
 $1$ &$-2$ &$+10$ &$-2$ &0&$+9$ &$+3$& $+792$\textbf{(+$\frac{44}{7}$)}& $\sigma_{23}$  \\[0.3em]
$T_{7}^0$ &   $\left(0^5~;
\frac{-1}{6}~\frac{-1}{6}~\frac{5}{6}\right)
 (0^5~\frac{-1}{4}~\frac{-1}{4}~\frac{1}{2})'$ & $\one_{0}$ & $2_{\bar 1}$ & $1$
&$-2$ &$-2$ &$+10$ &0&$+9$ &$+3$&$-1476$\textbf{(-$\frac{82}{7}$)}  & $\sigma_{24}$  \\[0.3em]
\hline

\end{tabular}
\end{center}
\caption{Left-handed \SUflip$\times$SU(5)$'\times$ SU(2)$'$ singlet states.  
$(N^L)_{j}$ is the notation for oscillator mode of the oscillating string with $j$ denoting the coordinate in the internal space, and ${\cal P}(f_{0,+,-})$ is the multiplicity of the corresponding spectrum in the twisted sector $T^{0,+,-}$. In this table, there is only ${\cal P}(f_0)$.
  The right-handed states in $T_3$ and $T_5$ are converted to the left-handed ones of $T_9$ and $T_7$,
respectively. }\label{tb:singlets} }
\end{table}

 \section{Mass scales}\label{sec:MassScale}
   
\begin{table}[!t]
{\tiny
\begin{center}
\begin{tabular}{|c|c|c|c||ccccccc|c|c|}
\hline Sect. & Charged singlet states & SU(5)$_X$  &
Mu. & $Q_1$& $Q_2$ & $Q_3$ & $Q_4$ & $Q_5$ & $Q_6$ &  $Q_{\rm anom}$ & La.&$Q_a^{\gamma\gamma}$  \\[0.3em]
\hline\hline

$U$ & $\left( +\,+\,+\,+\,+\,;-\,+\,-\,
\right)(0^8)'$ & $\one_{-5}$  &  &$-6$ &$+6$ &$-6$ &0 &0 &0 &$+630$\textbf{(+5)}  & $S_1$&$+630$  \\[0.5em]
\hline

$T_{4}^0$ &
$\left( +\,+\,+\,+\,+\,;\frac{-1}{6}\,
\frac{-1}{6}\,\frac{-1}{6}\,\right)
\left(0^8\right)'$ &  $\one_{-5}$ & $2$ &$-2$ &$-2$ &$-2$ &$0$ &$0$ &$0$ &$-378$\textbf{(-3)} & $S_{24}$&$-378$  \\[0.5em]
\hline

$T_{4}^+$ &
$\left((\frac{1}{6})^5\, ;\frac{-1}{6}\,
\frac{1}{6}\,\frac{1}{2}\,\right)
\left(\frac{1}{6}\,\frac{1}{6}\,\frac{1}{6}\,\frac{1}{6}\,\frac{-1}{2}\,;\frac{-1}{6}\,
\frac{-1}{2}\,\frac{1}{2}\,
\right)'$ &  $\one_{-5/3}$ & $2$ &$-2$ &+2 &+6 &$+4$ &$+10$ &$-10$ &$-666$\textbf{(-$\frac{37}{7}$)} & $S_2$&$-74$ \\[0.5em]
  &$\left((\frac{1}{6})^5\, ;\frac{-1}{6}\,
\frac{1}{6}\,\frac{1}{2}\,\right)
\left(\frac{1}{6}\,\frac{1}{6}\,\frac{1}{6}\,\frac{1}{6}\,\frac{-1}{2}\,;\frac{-1}{6}\,
\frac{1}{2}\,\frac{-1}{2}\,
\right)'$ & $\one_{-5/3}$ & 2 & $-2$& $+2$ &$+6$ & $-8$ & $-8$ & $-16$ &$+522$\textbf{(+$\frac{29}{7}$)}   & $S_3$   & $+58$\\[0.5em]
\hline

$T_{4}^-$ &
$\left(( \frac{-1}{6})^5\, ;\frac{-1}{6}\,
\frac{-1}{2}\,\frac{1}{6}\,\right)
\left((\frac{-1}{6})^4 \, \frac{1}{2}\,;\frac{1}{6}\,
\frac{+1}{2}\,\frac{-1}{2}\,
\right)'$ &  $\one_{5/3}$ & $2$ &--2 &--6 &+2 &$-4$ &$-10$ &$+10$ &$-594$\textbf{(-$\frac{33}{7}$)} & $S_4$&$-66$ \\[0.2em]
  &
$\left((\frac{-1}{6})^5\, ;\frac{-1}{6}\,
\frac{-1}{2}\,\frac{1}{6}\,\right)
\left((\frac{-1}{6})^4 \, \frac{1}{2}\,;\frac{1}{6}\,
\frac{-1}{2}\,\frac{+1}{2}\,
\right)'$ &$\one_{5/3}$&2  &$-2$ &$-6$  &$+2$  & $+8$ & $+8$ &$+16$ &$-1782$\textbf{(-$\frac{99}{7}$)}   & $S_5$ &$-198$  \\[0.5em]
\hline

$T_{2}^+$ &
$\left( (\frac{1}{3})^5\,;\frac{-1}{3}\,
\frac{1}{3}\,0\,\right)
\left((\frac{-1}{6})^4 \, \frac{1}{2}\,;\frac{-1}{3}\,
0\,\frac{1}{2}\,
\right)'$ &  $\one_{-10/3}$ & $1$ &$-4$ &$+4$  &$0$&$-4$ &$+8$ &$+16$  &$-396$\textbf{(-$\frac{22}{7}$)} & $S_6$&$-176$ \\[0.5em]
  &
$\left(  (\frac{-1}{6})^5\,;\frac{1}{6}\,
\frac{-1}{6}\,\frac{1}{2}\,\right)
\left((\frac{-1}{6})^4 \, \frac{1}{2}\,;\frac{2}{3}\,0\,
\frac{-1}{2}\,
\right)'$ &$\one_{5/3}$ &1  &$+2$ & $-2$ &$+6$ &$+8$  &$-10$  & $+10$ & $+162$\textbf{(+$\frac{9}{7}$)}  & $S_7$&$+18$   \\[0.5em]
  &
$\left( (\frac{-1}{6})^5\,;\frac{1}{6}\,
\frac{-1}{6}\,\frac{1}{2}\,\right)
\left((\frac{-1}{6})^4 \, \frac{1}{2}\,;\frac{-1}{3}\,0\,
\frac{1}{2}\,
\right)'$ &$\one_{5/3}$ &1 &$+2$ &$-2$&$+6$ &$-4$ &$+8$&$+16$ &$-1026$\textbf{(-$\frac{57}{7}$)}  & $S_8$&$-114$   \\[0.5em]
\hline

$T_{2}^-$ &
$\left( (\frac{-1}{3})^5\,;\frac{-1}{3}\,
0\,\frac{1}{3}\,\right)
\left(\frac{1}{6}\,\frac{1}{6}\,\frac{1}{6}\,\frac{1}{6}\,\frac{-1}{2}\,;\frac{1}{3}\,
 0\,\frac{-1}{2}\,
\right)'$ &  $\one_{10/3}$  & $1$ &$-4$ &$0$ &$+4$ &$+4$ &$-8$ &$-16$ &$+144$\textbf{(+$\frac{8}{7}$)} & $S_9$&$+64$ \\[0.5em]
  &
$\left( (\frac{1}{6})^5\,;\frac{1}{6}\,
\frac{-1}{2}\,\frac{-1}{6}\,\right)
\left(\frac{1}{6}\,\frac{1}{6}\,\frac{1}{6}\,\frac{1}{6}\,\frac{-1}{2}\,;\frac{-2}{3}\,0\,
\frac{1}{2}\,
\right)'$ &$\one_{-5/3}$ &1&$+2$ &$-6$ &$-2$&$-8$ & $+10$ &$-10$ &$-1170$\textbf{(-$\frac{65}{7}$)}  & $S_{10}$&$-130$ \\[0.5em]
  &
$\left( (\frac{1}{6})^5\,;\frac{1}{6}\,
\frac{-1}{2}\,\frac{-1}{6}\,\right)
\left(\frac{1}{6}\,\frac{1}{6}\,\frac{1}{6}\,\frac{1}{6}\,\frac{-1}{2}\,;\frac{1}{3}\,0\,
\frac{-1}{2}\,
\right)'$ &$\one_{-5/3}$ &1&$+2$ &$-6$  &$-2$  &$+4$ &$-8$  & $-16$&$+18$\textbf{(+$\frac{1}{7}$)}  & $S_{11}$&$+2$ \\[0.5em]
\hline

$T_{1}^+$ &
$\left( (\frac{-1}{3})^5\,;\frac{-1}{6}\,
\frac{1}{6}\,\frac{1}{2}\,\right)
\left(\frac{1}{6}\,\frac{1}{6}\,\frac{1}{6}\,\frac{1}{6}\,  \frac{-1}{2}\,;\frac{1}{12}\,\frac{-1}{4}\,0\,
\right)'$ &  $\one_{10/3}$ & $1$ &$-2$ &$+2$ &$+6$ &$+4$&$+1$ &$-13$ &$-72$\textbf{(-$\frac{4}{7}$)}  & $S_{12}$&$-32 $ \\[0.5em]
  &
$\left( (\frac{1}{6})^5\,;\frac{-2}{3}\,
\frac{2}{3}\,0\,\right)
\left(\frac{1}{6}\,\frac{1}{6}\,\frac{1}{6}\,\frac{1}{6}\,\frac{-1}{2}\,;\frac{1}{12}\,
\frac{-1}{4}\,0\,
\right)'$ &$\one_{-5/3}$ &1  &$-8$ &$+8$&0 &$+4$&$+1$& $-13$&$+558$\textbf{(+$\frac{31}{7}$)}  & $S_{13}$&$+62$   \\[0.5em]
  &
$\left( (\frac{1}{6})^5\,;\frac{1}{3}\,
\frac{-1}{3}\,0\,\right)
\left(\frac{1}{6}\,\frac{1}{6}\,\frac{1}{6}\,\frac{1}{6}\,\frac{-1}{2}\,;\frac{1}{12}\,
\frac{-1}{4}\,0\,
\right)'$ &$\one_{-5/3}$ &2 &$+4$ &$-4$&0& $+4$&$+1$&$-13$&$-198$\textbf{(-$\frac{11}{7}$)}  & $2S_{14}$&$-22$   \\[0.5em]
\hline

$T_{1}^-$ &
$\left( (\frac{1}{3})^5\,;\frac{-1}{6}\,
\frac{1}{2}\,\frac{1}{6}\,\right)
\left((\frac{-1}{6})^4 \, \frac{1}{2}\,;
\frac{5}{12}\,\frac{-1}{4}\,0\,
\right)'$ &  $\one_{-10/3}$ & $1$ &$-2$ &$+6$&$+2$& $+8$&$-1$ &$+13$&$+576$\textbf{(+$\frac{32}{7}$)}  & $S_{15}$&$+256$ \\[0.5em]
  &
$\left( (\frac{-1}{6})^5\,;\frac{-2}{3}\,
0\,\frac{-1}{3}\, \right)
\left((\frac{-1}{6})^4 \, \frac{1}{2}\,;\frac{5}{12}\,
\frac{-1}{4}\,0\,
\right)'$ &$\one_{5/3}$ &1  &$-8$&0&$-4$&$+8$&$-1$&$+13$&$-558
$\textbf{(-$\frac{31}{7}$)}  & $S_{16}$&$-62$   \\[0.5em]
  &
$\left( (\frac{-1}{6})^5\,;\frac{1}{3}\,
0\,\frac{2}{3}\,\right)
\left((\frac{-1}{6})^4 \, \frac{1}{2}\,;\frac{5}{12}\,
\frac{-1}{4}\,0\,
\right)'$ &$\one_{5/3}$ &1 &$+4$ &0& $+8$&$+8$&$-1$&$+13$&$-54$\textbf{(-$
\frac{3}{7}$)}  & $S_{17}$&$-6$   \\[0.5em]
\hline

$T_{7}^+$ &
$\left( (\frac{-1}{3})^5\,;\frac{-1}{6}\,
\frac{1}{6}\,\frac{-1}{2}\,\right)
\left(\frac{1}{6}\,\frac{1}{6}\,\frac{1}{6}\,\frac{1}{6}\,\frac{-1}{2}\,;\frac{-5}{12}\,
\frac{1}{4}\,0\,
\right)'$ &  $\one_{-10/3}$  & $1$ &$-2$ &$+2$ &$-6$ & $-8$&$+1$&$-13$& $+432$\textbf{(+$\frac{24}{7}$)} & $S_{18}$&$+192$ \\[0.5em]
  &
$\left( (\frac{1}{6})^5\,;\frac{1}{3}\,
\frac{2}{3}\,0\,\right)
\left(\frac{1}{6}\,\frac{1}{6}\,\frac{1}{6}\,\frac{1}{6}\,\frac{-1}{2}\,;\frac{-5}{12}\,
\frac{1}{4}\,0\,
\right)'$ & $\one_{5/3}$&1 &$+4$ &$+8$ &0& $-8$&$+1$ &$-13$ &$+1566$\textbf{(+$\frac{87}{7}$)}  &$S_{19}$ &$+174$   \\[0.5em]
  &
$\left( (\frac{1}{6})^5\,\,;\frac{-2}{3}\,
\frac{-1}{3}\,0\,\right)
\left(\frac{1}{6}\,\frac{1}{6}\,\frac{1}{6}\,\frac{1}{6}\,\frac{-1}{2}\,;\frac{-5}{12}\,
\frac{1}{4}\,0\,
\right)'$ & $\one_{5/3}$&1 &$-8$ &$-4$ &0&$-8$&$+1$ &$-13$ &$-1206$\textbf{(-$\frac{67}{7}$)}  &  $S_{20}$&$-134$ \\[0.5em]
\hline

$T_{7}^-$ &
$\left( (\frac{1}{3})^5\,;\frac{-1}{6}\,
\frac{-1}{2}\,\frac{1}{6}\,\right)
\left((\frac{-1}{6})^4 \, ,\frac{1}{2}\,;
\frac{-1}{12}\,\frac{1}{4}\,0\,
\right)'$ &  $\one_{10/3}$  & $1$ &$-2$&$-6$& $+2$& $-4$& $-1$& $+13$& $-1188$\textbf{(-$\frac{66}{7}$)} & $S_{21}$&$-528$ \\[0.5em]
 &$\left( (\frac{-1}{6})^5\,;\frac{-2}{3}\,0\,
\frac{2}{3}\,\right)
\left((\frac{-1}{6})^4 \,\frac{1}{2}\,;
\frac{-1}{12}\,\frac{1}{4}\,0\,
\right)'$  &$\one_{-5/3}$ &1 & $-8$&0&$+8$&$-4$ &$-1$ &$+13$ &$-1062$\textbf{(-$\frac{59}{7}$)}  & $S_{22}$&$-118$ \\[0.5em]
  &
$\left( (\frac{-1}{6})^5\,;\frac{1}{3}\,0\,
\frac{-1}{3}\,\right)
\left((\frac{-1}{6})^4 \, \frac{1}{2}\,;
\frac{-1}{12}\,\frac{1}{4}\,0\,
\right)'$  &$\one_{-5/3}$ &2 & $+4$&0&$-4$ &$-4$ &$-1$& $+13$&$+450$\textbf{(+$\frac{25}{7}$)}  & $S_{23}$& $+100$ \\[0.5em]
\hline\hline
& &  & &$-16$ &$-28$  &$+8$ &0&$+18$ &$+42$ &$-7632\qquad~~$ & &$-1162$   \\[0.3em]
\hline

\end{tabular}
\end{center}
\caption{Electromagnetically charged singlets.  }\label{tb:ChSinglets}
}
\end{table}

 Below the Planck scale $\Mp$, we consider four scales, the compactification scale allowing large masses $\Mv$ to vector-like pairs, the GUT scale \Mgt, the intermediate \Mit, and  the electroweak scale \vew.
The principle of removing vector-like pairs is just the gauge principle as emphasized in \cite{KimPRD17}. If extra symmetries are introduced, one must include another assumption(s) how those extra
symmetries are broken. 
The hierarchy of scales that we consider is
 \begin{eqnarray}
&&{\rm E}_8\times  {\rm E}_8'{\quad\longrightarrow\quad} {\rm GUT}{\quad\longrightarrow\quad} {\rm SM~ and ~``invisible''~axion} {\quad\longrightarrow\quad} {\rm SU(3)_c}\times{\rm U(1)_{em}}\nonumber\\
&&\qquad\quad\Mv\qquad\quad\Mg\qquad\quad\Mi\qquad
\qquad\qquad\qquad\qquad v_{\rm ew} \nonumber
 \end{eqnarray}
 where $\Mv^2$ is the order of the  string tension, $\alpha^{\prime\,-1}$.
 The particles removed at the compactification scale are the vector-like sets. In Table \ref{tb:ChSinglets}, we list charged singlets. The vector-like sets,  including the charges $\Qanom$ and $\Z_{12}$ orbifolds, must be removed at \Mvt. 
  The U(1)'s which can be broken at the GUT scales are five anomaly free $\tilde{\rm U}(1)$'s. \Uanom~global symmetry is required to be broken at the axion window and works as a global symmetry at the GUT scale. Any singlet of Table \ref{tb:ChSinglets} can have a GUT scale VEV, leaving a global symmetry below $\Mg$ via the 't Hooft mechanism \cite{Hooft71}. We can repeat this process for breaking all five anomaly free $\tilde{\rm U}(1)$'s, leaving only \Uanom~global symmetry below $\Mg$.
So, the doublet-triplet splitting can be of the form ``colored particles=GUT scale and $H_{u,d}$=light''. Even if the masses of colored particles are a bit smaller than $M_{GUT}$, proton stability can be achieved.  Proton decay by dimension 6 operators of quarks and leptons (by the exchange of colored scalars)  is helped by  the  Yukawa couplings for the first family membersby the order $10^{-5.5}$. Then, it is not problematic. With SUSY assumption, dimension 5 operators of quark and lepton superfields, $W_4\propto \overline{\bf 10}_{-1}\cdot\overline{\bf 10}_{-1}\cdot\overline{\bf 10}_{-1}  \cdot {\bf 5}_{+3}$, is the leading contribution. This is disastrously dangerous if the colored scalar masses are somewhat smaller than  $\Mg$.  There can arise a $\Z_4$  from a subgroup of an anomaly free $\tilde{\rm U}(1)$ gauge group, eliminating  $W_4$  as shown  in \cite{KimPLB13} and there is no fast proton decay problem \cite{Leeetal11}. 
  
 There are two GUT scale sets,
 \begin{eqnarray}
 C_7^{T_6}+C_8^{T_6},~C_{11}^{T_3}+C_{12}^{T_9}.
 \end{eqnarray}
These are \SUflip~non-singlets and these vector-like sets cannot be external light fields. In addition, the superpotential terms cannot be generated through the intermediate states of these vector-like fields. Diagrams with these internal \Mgt~mass states must contain loops, which cannot generate superpotential terms because of the non-renomalization theorem.  So, high dimension superpotential terms composed of light fields cannot be generated with a suppression factor $\Mg$  or \Mvt~ in our model. 
   
 If a high dimension superpotential term of light fields is generated, the relevant mass suppression factor must be the intermediate scale \Mit. This conclusion, depending on our detailed model, is very different from the general strategy in the MSSM where the $\mu$ term for example has the suppression factor $\Mg$ or   $\Mp$ \cite{KimNilles84}.
 
\section{Yukawa couplings}  \label{sec:Yukawa}  
      
Now, we search for a possibility for
a non-zero (33) element of $M_{\rm Higgsino}^{3\times 3}$ in Eq. (\ref{eq:Mhigginos}), which respects the \Uanom~symmetry and $\Z_{12-I}$ selection rules.  At dimension 10 level, there appears one,
\begin{eqnarray}
 W\propto   \frac{1}{\Mi^7} H_u^{(T_7^0)}H_d^{(T_7^0)}\,\sigma_{12}^{(T3)}
  \sigma_{23}^{(T_7^0)}\,\sigma_{21}^{(T_7^0)}\sigma_{21}^{(T_7^0)} \,\sigma_{8}^{(T6)}\sigma_{8}^{(T6)}
  \sigma_{8}^{(T6)}\sigma_{1}^{(T_4^0)}
  \end{eqnarray}
  where we used \Mit~as the suppression factor.
This term introduces a nonzero entry $\varepsilon$ in the (33) element which cannot be very small because the VEVs of singlets are also at the scale \Mit. In this case, the remaining two eigenstates of (\ref{eq:Mhigginos}) are
 \begin{eqnarray}
\frac{1}{N}\begin{pmatrix}
1\\ 1\\[0.5em] -\frac{M}{m}+\frac{\varepsilon}{2m}-\sqrt{2+\left(\frac{M}{m}-\frac{\varepsilon}{2m}\right)^2}
  \end{pmatrix}  ,~~{\rm mass}=M+\frac{\varepsilon}{2}-\sqrt{2m^2+\left(M-\frac{\varepsilon}{2}\right)^2}
 \end{eqnarray}
and the massless one is still Eq. (\ref{eq:Massless}),
 \begin{eqnarray}
{\rm \five} ^{T_4^0},{\rm \fiveb} ^{T_4^0}=\frac{1}{\sqrt2}\begin{pmatrix}1\\ -1\\ 0\end{pmatrix}, ~{\rm mass}=0 .\label{eq:HiggsQuint}
\end{eqnarray}
At this level, the Higgs doublets are those appearing in $T_4^0$.
However, the doublets from  $T_4^0$ obtain mass at the electroweak scale when soft masses of order $m_{3/2}^2$ are introduced, which is a well-known fact in the supergravity phenomenology, which will not be discussed here.

\subsection{Doublet-triplet splitting}\label{subsec:DT}

\begin{table}[!t]
{\tiny
\begin{center}
\begin{tabular}{|c|c|c |c||ccccccc|c|}
\hline Sect. & $P+kV$ & SU(5)$_X$ &
Mult.& $Q_1$& $Q_2$ & $Q_3$ & $Q_4$ & $Q_5$ & $Q_6$ &  $Q_{\rm anom}$ & Label   \\[0.5em]
\hline\hline
 
$T_{8}^0$ &
$\left(\underline{+\,+\,-\,-\,-\,};\frac{+1}{6}\,\frac{+1}{6}\,\frac{+1}{6}\,
\right)(0^8)'$ &  $\ten_{+1}$ & $2$ &$+2$ &$+2$ &$+2$ &0 &0 &0 &$+378 {\bf (\frac{+21}{7})}$ & $ C_{\rm v1}$  \\[0.5em]
$T_{4}^0$ &
$\left(\underline{+\,+\,+\,-\,-\,};\frac{-1}{6}\,\frac{-1}{6}\,\frac{-1}{6}\,
\right)(0^8)'$ &  $\tenb_{-1}$ & $2$ &$-2$ &$-2$ &$-2$ &0 &0 &0 &$-378  {\bf (\frac{-21}{7})}$ & $ C_{\rm v2}$  \\[0.5em]
$T_{5}^0$ &
$\left(\underline{-1\,0\, 0\,0\,0\,};\frac{+1}{6}\,\frac{+1}{6}\,\frac{ +1}{6}\,
\right)(0^5;\frac{+1}{4}, \frac{+1}{4} ,\frac{-2}{4} )'$ &   $\fiveb_{+2}$ & $1$ &$+2$ &$+2$ &$+2$ &0 &$-9$ &$-3$ &$+972  {\bf (\frac{+54}{7})}$ & $  C_{\rm v3}$  
 \\[0.3em]
\hline

\end{tabular}
\end{center}
\caption{Three left-handed states belonging to the vector-like spectra appearing in Fig. \ref{fig:HighDim}. }\label{tb:Vectorlike} }
\end{table}

\begin{figure}[!t]
\begin{center}
\includegraphics[width=0.6\textwidth]{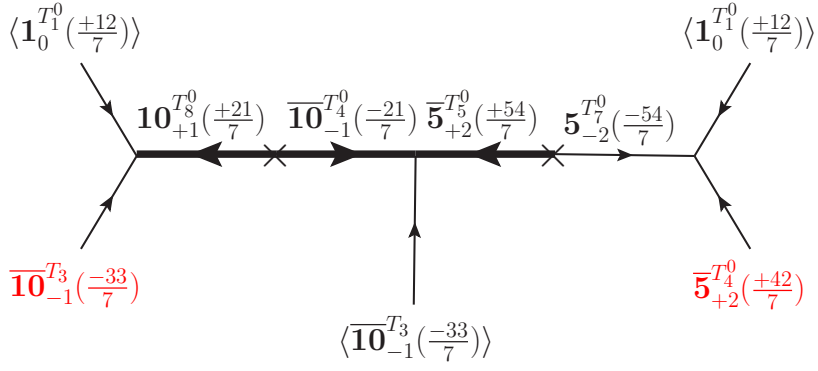} 
\end{center}
\caption{ A high dimensional term. The fractional numbers in the brackets are $\Qanom/126$.} \label{fig:HighDim}
\end{figure}

For the successful MSSM, the multiplet
(\ref{eq:HiggsQuint}) must be split into heavy colored ones and light Higgs doublets.  This doublet-triplet splitting problem is achieved by the VEVs of $\tenb_{-1}$ and $\ten_{+1}$ appearing in  $T_3$ and $T_9$. Here, we show it explicitly from a detailed string compactification model discussed above. Here we show that the suppression factor is the   mass of vector-like pair $\Mv$. The VEVs   of ${\tenb}_{-1}$ and $ {\ten}_{+1}$ give mass to the colored triplets of the Higgs quintets, ${\rm \five}_{-2} ^{T_4^0}$ and ${\rm \fiveb}^{T_4^0}_{+2}$.  But the problem is the scale for the effective operator. We find the following operators,
\begin{eqnarray}
&&{\rm VEV~of~\tenb}:~\frac{1}{\Mv\Mi} \, \langle C_{11}^{T3}\rangle\langle\sigma_{21}^{T_1^0}  \rangle  \langle\sigma_{21}^{T_1^0}  \rangle \, C_{11}^{T3}\,C_6^{T_4^0} \to d_{\rm from\, {\tenb}_{-1}}\,d^c_{\rm from\, {\fiveb}_{+2}}  ,
\\[0.5em]
 &&{\rm VEV~of~\ten}:~\frac{1}{\Mv \Mi} \,  \langle  C_{12}^{T9}\rangle\langle\sigma_{19}^{T_1^0}  \rangle  \langle\sigma_{20}^{T_1^0}  \rangle \, C_{12}^{T9}\,C_5^{T_4^0} \to d^c_{\rm from\, {\ten}_{+1}}\, 
 d_{\rm from\, {\five}_{-2}},
 \end{eqnarray}
 where $1/\Mv$ suppression is because there appear heavy vector-like states in the tree diagram.\footnote{Loop diagrams are not considered because of the non-renormalization theorem.} Two thick lines of Fig. \ref{fig:HighDim} are some vector-like states of Ref. \cite{Huh09} because there is no massless (left-handed) states in $T_8^0$ and $T_5^0$. The cross in the RHS is of order $\Mi$ because  ${\one}_0^{T_7^0}(\frac{-54}{7})$ is present in Table \ref{tb:singlets}. On the other hand, the cross in the LHS is expected of order $\Mv$ because ${\tenb}^{T_4^0}(\frac{-21}{7})$ does not appear  in Table \ref{tb:singlets}. Even though the left-handed states of Table \ref{tb:singlets} do not contain a state with $\Qanom=\frac{-21}{7}$, the vector-like pairs ($\ten$ and $\tenb$ shown as two thick arrows) can fulfill the quantum numbers because we will not require the masslessness conditions in the orbifold selection rules for the vector-like states. Also, the thick arrow line for $\fiveb_{+2}$ does not appear in  Table \ref{tb:singlets}.  Here, these three  states are denoted as `vector-like states' which are shown in Table \ref{tb:Vectorlike}. The left-mover and right-mover masses in the heterotic string with $\Z_{12-I}$ compactification in the $k^{\rm th}$ twisted sector are given by,
 \begin{eqnarray}
 &&M_L^2= \frac{(P+kV)^2}{2}-\frac{2\tilde{c}_k}{2}+\frac{2\tilde{N}_L}{2},\\[0.5em]
 && M_R^2= \frac{(\tilde{s}+ k\phi)^2}{2} -\frac{2c_k }{2}+\frac{2\tilde{N}_R}{2},
 \end{eqnarray}
where $\tilde{N}_{L,R}$ are the oscillator numbers and $2\tilde{c}_{4,8}=\frac32$ and $2{c}_{4,8}=\frac12$ are given in Ref.  \cite{KK07}. In the model of \cite{KK07}, the L and R vectors are
 \begin{eqnarray}
  &&V=(0^5;\frac{-1}{6},\frac{-1}{6},\frac{-1}{6})(0^5;\frac{+1}{4},\frac{+1}{4},\frac{-2}{4})',\\[0.5em]
&&\phi=(\frac{5}{12},\frac{4}{12},\frac{1}{12}).
 \end{eqnarray}

 The  QCD-color field  in ${\fiveb}_{+2}$ finds  the partner in ${\tenb}_{-1}$, both of which are shown in reds in Fig. \ref{fig:HighDim}. Similarly,  ${\five}_{-2}$   finds  the partner in  ${\ten}_{+1}$.  Thus, there remain only massless Higgs doublets from Eq. (\ref{eq:HiggsQuint}), $H_u$ and $H_d$, and the doublet-triplet splitting is realized. One pair of   ${\five}_{-2}$ and ${\fiveb}_{+2}$ needs one pair of   ${\ten}_{+1}$  and ${\tenb}_{-1}$ for the splitting.
 
 In  view of the longevity of proton, note that the dangerous $ \tenb_H \cdot \tenb\cdot\tenb\cdot\five$ and $\tenb_H\cdot \five\cdot\five\cdot\one$ couplings are allowed in \SUflip, judging from the gauge quantum numbers alone. In the ordinary SO(10) which is the covering group of \SUflip, these terms are forbidden by imposing the R-parity quantum numbers --1 and +1, respectively, for the matter and Higgs fields.  In our case, the GUT scale Higgs boson $\tenb_H$ and $\ten_H$ are $C_{11}$ and $C_{12}$ of Table \ref{tb:colorfields}. These are from $T_3^0$ and $T_9^0$. The orbifold selection rules allow the superpotential term $ C_{11}C_{12}$ but does not allow $C_{11} C_4C_4 C_3$ and $C_{11} C_3C_3\cdot \one$ from the fields in Table 1. In the latter operator containing the $L$-violating $LLE^c$, $\one$ may be chosen from $T_1^0$ such as $\sigma_{19,20,21}$ of Table \ref{tb:singlets}. But, $LLE^c$ alone does not trigger proton decay. Basically, the R-parity interpretation for the proton longevity in SO(10) GUT, up to dimension 5 operators, is automatic from our orbifold selection rules \cite{CK06} in case of the \SUflip. But, a complete study on the proton longevity is outside the scope of this paper.
  
 \subsection{The CKM and PMNS matrices}
 
\begin{table}[!t]
{\tiny
\begin{center}
\begin{tabular}{|c|c|c|c||ccccccc|c|c|}
\hline & $P+kV$ & SU(5)$'$ & Mu. & $Q_1$& $Q_2$ & $Q_3$ & $Q_4$ & $Q_5$ & $Q_6$ &  $Q_{\rm anom}$ & La. &$Q_a^{\gamma\gamma}$  \\[0.3em]
\hline\hline

$T_{1}^0$ &
$\left(0^5;\frac{-1}{6}\,\frac{-1}{6}\,\frac{-1}{6}\,\right)
(\underline{-1\,0^3}\,0\,; \frac{1}{4}\,\frac{1}{4}\,\frac{1}{2} )'$ &  $\tenb'_0$  & $1$ &$-2$ &$-2$ &$-2$ &0 &$+3$ & $+9$ &$-648$\textbf{(-$\frac{36}{7}$)}  & $T_1'$&0   \\[0.5em]
 & $\left(0^5;\frac{-1}{6}\,\frac{-1}{6}\,\frac{-1}{6}\,\right)
(\underline{\frac{1}{2}\,\frac{1}{2}\,\frac{-1}{2}\,\frac{-1}{2}\,}\,\frac{1}{2}\,; \frac{-1}{4}\,\frac{-1}{4}\,0 )'$ &      &&&&&&&& & & \\[0.5em]
\hline
$T_{1}^0$ &$\left(0^5\,;\frac{-1}{6}\,\frac{-1}{6}\,\frac{-1}{6}\,\right)
(\underline{1\,0\,0\,0\,}\,0\,; \frac{1}{4}\,\frac{1}{4}\,\frac{1}{2} )'$ &$(\five',\two')_0$& $1$ &$-2$ &--2 &--2 &0  &$+3$ &$-3$ &$-540$\textbf{(-$\frac{30}{7}$)}  & $F_1'$&0 \\[0.2em]
&$\left(0^5\,;\frac{-1}{6}\,\frac{-1}{6}\,\frac{-1}{6}\,\right)
({0\,0\,0\,0\,}\,0\,; \frac{-3}{4}\,\frac{-3}{4}\,\frac{-1}{2} )'$ &  &&&&&&&& &&   \\[0.2em]
&$\left(0^5\,;\frac{-1}{6}\,\frac{-1}{6}\,\frac{-1}{6}\,\right)
(\underline{\frac12\,\frac{-1}{2}\,\frac{-1}{2}\,\frac{-1}{2}\,}\,\frac{-1}{2}\,\,; \frac{-1}{4}\,\frac{-1}{4}\,0 )'$ &  &&&&&&&& &&   \\[0.2em]
&$\left(0^5\,;\frac{-1}{6}\,\frac{-1}{6}\,\frac{-1}{6}\,\right)
({\frac12\,\frac{1}{2}\,\frac{1}{2}\,\frac{1}{2}\,}\,\frac{-1}{2}\,\,; \frac{-1}{4}\,\frac{-1}{4}\,0 )'$ &  &&&&&&&& &&   \\[0.6em]
\hline
$T_{1}^0$&$\left(0^5\,;\frac{-1}{6}\,\frac{-1}{6}\,\frac{-1}{6}\,\right)
(\underline{\frac{-1}{2}\,\frac{1}{2}\,\frac{1}{2}\,\frac{1}{2}\,}\,\frac{-1}{2}\,\,; \frac{-1}{4}\,\frac{-1}{4}\,0 )'$ & $\fiveb'_0$  & $1$ &--2 &--2 &--2 &0 &$+3$ &$-15$ &$-432$\textbf{(-$\frac{24}{7}$)}  &  $F_2'$ &0  \\[0.2em]
&$\left(0^5\,;\frac{-1}{6}\,\frac{-1}{6}\,\frac{-1}{6}\,\right)
(0\,0\,0\,0\,-1 \,; \frac{1}{4}\,\frac{1}{4}\,\frac{1}{2} )'$ && &&&&&&&&& \\[0.3em]
\hline

$T_{1}^+$
&$\left((\frac{1}{6})^5 \,;\frac{1}{3}\,
\frac{-1}{3}\,0 \,\right)
\left(\underline{\frac{-5}{6}\,\frac{1}{6}\,\frac{1}{6}\,\frac{1}{6}\,}\,\frac{1}{2}\,
;\frac{1}{12}\,\frac{-1}{4}\,0\,\right)'$ &  $\fiveb'_{-5/3}$   & $1$ &$+4$ &$-4$ &0 &$+4$
&$+1$ &$+11$ &$-414$\textbf{(-$\frac{23}{7}$)}  & $F'_3$&$-230$ \\[0.2em]
&$\left( (\frac{1}{6})^5 \,;\frac{1}{3}\,
\frac{-1}{3}\,0 \,\right)
\left(\frac{-1}{3}\,\frac{-1}{3}\,\frac{-1}{3}\,\frac{-1}{3}\,0\,;\frac{7}{12}\,
\frac{1}{4}\,\frac{1}{2}\,\right)'$ &&&&&&&&& &&  \\[0.3em]
\hline

$T_{4}^+$ &
$\left((\frac{1}{6})^5 \,;\frac{-1}{6}\,
\frac{1}{6}\,\frac{1}{2}\,\right)
\left(\underline{\frac{2}{3}\,\frac{-1}{3}\,\frac{-1}{3}\,\frac{-1}{3}\,}\,0\,;
\frac{1}{3}\,0\,0\,\right)'$ &  $\five'_{-5/3}$  & $3$ &--2 &+2 &+6 &$+4$ &$-2$ &$+2$ &$-18$\textbf{(-$\frac{1}{7}$)}  & $F'_4$&$-10$ \\[0.2em]
 &$\left((\frac{1}{6})^5  \,;
 \frac{-1}{6}\,\frac{1}{6}\,\frac{1}{2}\,\right)
\left(\frac{1}{6}\,\frac{1}{6}\,\frac{1}{6}\,\frac{1}{6}\,\frac{1}{2}\,;
\frac{-1}{6}\,\frac{-1}{2}\,\frac{-1}{2}\,\right)'$ & && &&&&&& & &   \\[0.2em]
\hline
$T_{4}^-$ &
$\left((\frac{-1}{6})^5 \,;
\frac{-1}{6}\,\frac{-1}{2}\,\frac{1}{6}\,\right)
\left(\underline{\frac{-2}{3}\,\frac{1}{3}\,\frac{1}{3}\,\frac{1}{3}\,}\,0\,;
\frac{-1}{3}\,0\,0\,\right)'$ &  $\fiveb'_{5/3}$   & $3$ &--2 &--6 &+2 &$-4$
&$+2$ &$-2$ &$-1242$\textbf{(-$\frac{69}{7}$)} & $F'_5$&$-690$ \\
 &$\left((\frac{-1}{6})^5 \,;
 \frac{-1}{6}\,\frac{-1}{2}\,\frac{1}{6}\,\right)
\left((\frac{-1}{6})^4 \frac{-1}{2}\,;
\frac{1}{6}\,\frac{1}{2}\,\frac{1}{2}\,\right)'$ & && &&&&&& & &  \\[0.2em]
\hline

$T_{7}^-$
&$\left((\frac{-1}{6})^5 \,;
\frac{1}{3}\,0 \,\frac{-1}{3}\,\right)
\left(\underline{\frac{5}{6}\,\frac{-1}{6}\,\frac{-1}{6}\,\frac{-1}{6}\,}\,\frac{-1}{2}\,;
\frac{-1}{12}\,\frac{1}{4}\,0\,\right)'$ &  $\five'_{5/3}$ & $1$ &+4 &0 &--4 &$-4$
&$-1$ &$-11$  &$+666$\textbf{(+$\frac{37}{7}$)} & $F'_6$&$+370$ \\[0.2em]
&$\left((\frac{-1}{6})^5 \,;
\frac{1}{3}\,0 \,\frac{-1}{3}\,\right)
\left(\frac{1}{3}\,\frac{1}{3}\,\frac{1}{3}\,\frac{1}{3}\,0\,;\frac{-7}{12}\,
\frac{-1}{4}\,\frac{-1}{2}\,\right)'$ & &&&&&&&& &&  \\[0.3em]
\hline\hline
&   &&  &$-16$ &$-28$  &$+8$ &0 &$+18$ &$+6$&$-3492\qquad\quad$ &  &$ -1960$   \\[0.2em]
\hline

\end{tabular}
\end{center}
\caption{The SU(5)$'$ representations.  Notations are the same as in Table \ref{tb:colorfields}.  }\label{tb:SUhfields} }
\end{table}
\begin{table}[!t]
{\tiny
\begin{center}
\begin{tabular}{|c|c|c|c||ccccccc|c|c|}
\hline Sect. & $P+kV$ & SU(2)$'$  & Mult. & $Q_1$& $Q_2$ & $Q_3$ & $Q_4$ & $Q_5$ & $Q_6$ &  $Q_{\rm anom}$ & Label&$Q_a^{\gamma\gamma}$  \\[0.3em]
\hline\hline

$T_1^0$
&$\left(0^5\,;\frac{-1}{6}\,\frac{-1}{6}\,\frac{-1}{6}\,\right)
(\underline{1\,0\,0\,0\,}\,0\,; \frac{1}{4}\,\frac{1}{4}\,\frac{1}{2} )'$ &  $(\five',\two')_0$   & $1$ &--2 &--2 &--2 &0 &$+3$ &$-3$ &$-540$\textbf{(-$\frac{30}{7}$)} & $ D_1'$ &In\\[0.2em]
&$\left(0^5\,;\frac{-1}{6}\,\frac{-1}{6}\,\frac{-1}{6}\,\right)
({0\,0\,0\,0\,}\,0\,; \frac{-3}{4}\,\frac{-3}{4}\,\frac{-1}{2} )'$ &&&&&&&&& && Ta. \ref{tb:SUhfields}   \\[0.5em]
\hline

$T_{1}^0$ &
$\left(0^5\,;\frac{-1}{6}\,
\frac{-1}{6}\,\frac{-1}{6}\,\right)
\left(0\,0\,0\,0\,1\,;\frac{1}{4}\,\frac{1}{4}\,\frac{1}{2}\,
\right)'$ &  $\two'_0$  & $1$ &$-2$ &$-2$ &$-2$ &0 &$+3$ &$+21$ &$-756$\textbf{(-$6$)} & $D_2$&0 \\[0.5em]
\hline

$T_{1}^+$ &
$\left((\frac{1}{6})^5 \,;\frac{1}{3}\,
\frac{-1}{3}\,0\,\right)
\left(\frac{1}{6}\,\frac{1}{6}\,\frac{1}{6}\,\frac{1}{6}\,\frac{1}{2}\,;\frac{1}{12}\,
\frac{3}{4}\,0\,
\right)'$
&$\two'_{-5/3}$  & $1$ &$+4$ &$-4$ &$0$ &$-8$ &$-5$ &$+5$ &$+18$\textbf{(+$\frac{1}{7}$)} & $D_3$&$+4$ \\[0.5em]
\hline

$T_{1}^-$ &
$\left((\frac{-1}{6})^5\, ;\frac{-2}{3}\,
0\,\frac{-1}{3}\,\right)
\left(\frac{1}{3}\,\frac{1}{3}\,\frac{1}{3}\,\frac{1}{3}\,0\,;\frac{-1}{12}\,
\frac{1}{4}\,\frac{1}{2}\,\right)'$
&$\two'_{5/3}$  & $1$ &$-8$ &$0$ &$-4$ &$-4$ &$+5$ &$-5$ &$-774$\textbf{(-$\frac{43}{7}$)}  & $D_4$& $-172$ \\[0.5em]
\hline

$T_{1}^-$ &
$\left((\frac{-1}{6})^5 \,;\frac{1}{3}\,
0\,\frac{2}{3}\,\right)
\left(\frac{1}{3}\,\frac{1}{3}\,\frac{1}{3}\,\frac{1}{3}\,0\,;\frac{-1}{12}\,
\frac{1}{4}\,\frac{1}{2}\,\right)'$
&$\two'_{5/3}$  & $1$ &$+4$ &$0$ &$+8$ &$-4$ &$+5$ &$-5$ &$-270$\textbf{(-$\frac{15}{7}$)}  & $D_5$& $-60$ \\[0.5em]
\hline

$T_{2}^+$ &
$\left((\frac{-1}{6})^5 \,;\frac{1}{6}\,
\frac{-1}{6}\,\frac{1}{2}\,\right)
\left(\frac{1}{3}\,\frac{1}{3}\,\frac{1}{3}\,\frac{1}{3}\,0\,;\frac{1}{6}\,\frac{1}{2}\,0\,
\right)'$ &  $\two'_{5/3}$ & $1$ &$+2$ &$-2$  &$+6$&$-4$ &$-4$ &$-8$  &$-54$\textbf{(-$\frac{3}{7}$)} & $D_6$& $-12$  \\[0.5em]
\hline

$T_{2}^-$ &
$\left((\frac{1}{6})^5 \,;\frac{1}{6}\,
\frac{-1}{2}\,\frac{-1}{6}\,\right)
\left(\frac{1}{6}\,\frac{1}{6}\,\frac{1}{6}\,\frac{1}{6}\,\frac{1}{2}\,;\frac{1}{3}\,0\,
\frac{1}{2}\,
\right)'$ &  $\two'_{-5/3}$  & $1$ &$+2$ &$-6$ &$-2$ &$+4$ &$+4$ &$+8$ &$-954$\textbf{(-$\frac{53}{7}$)} & $D_7$& $-212$  \\[0.5em]
\hline

$T_{4}^+$ &
$\left((\frac{1}{6})^5 \,;\frac{-1}{6}\,
\frac{1}{6}\,\frac{1}{2}\,\right)
\left(\frac{1}{6}\,\frac{1}{6}\,\frac{1}{6}\,\frac{1}{6}\,\frac{1}{2}\,;\frac{-1}{6}\,
\frac{1}{2}\,\frac{1}{2}\,
\right)'$ &  $\two'_{-5/3}$ & $2$ &--2 &+2 &+6 &$-8$ &$+4$ &$+8$ &$-450$\textbf{(-$\frac{25}{7}$)}  & $ D_8$& $-100$ \\[0.5em]
\hline

$T_{4}^-$ &
$\left((\frac{-1}{6})^5 \,;\frac{-1}{6}\,
\frac{-1}{2}\,\frac{1}{6}\,\right)
\left(\frac{1}{3}\,\frac{1}{3}\,\frac{1}{3}\,\frac{1}{3}\,0\,;\frac{2}{3}\,0\,0\,
\right)'$ &  $\two'_{5/3}$ & $2$ &--2 &--6 &+2 &$+8$ &$-4$ &$-8$ &$-810$\textbf{(-$\frac{45}{7}$)} & $ D_9$& $-180$ \\[0.5em]
\hline

$T_{7}^+$ &
$\left((\frac{1}{6})^5 \,;\frac{1}{3}\,
\frac{2}{3}\,0\,\right)
\left(\frac{1}{6}\,\frac{1}{6}\,\frac{1}{6}\,\frac{1}{6}\,\frac{1}{2}\,;\frac{7}{12}\,
\frac{1}{4}\,0\,
\right)'$ &  $\two'_{5/3}$  & $1$ &$+4$ &$+8$ &$0$ &$+4$ &$-5$ &$+5$ &$+1782$\textbf{(+$\frac{99}{7}$)} & $D_{10}$& $+396$ \\[0.5em]
\hline

$T_{7}^+$ &
$\left((\frac{1}{6})^5 \,;\frac{-2}{3}\,
\frac{-1}{3}\,0\,\right)
\left(\frac{1}{6}\,\frac{1}{6}\,\frac{1}{6}\,\frac{1}{6}\,\frac{1}{2}\,;\frac{7}{12}\,
\frac{1}{4}\,0\,
\right)'$ &  $\two'_{5/3}$  & $1$ &$-8$ &$-4$ &$0$ &$+4$ &$-5$ &$+5$ &$-990$\textbf{(-$\frac{55}{7}$)} & $D_{11}$& $-220$ \\[0.5em]
\hline

$T_{7}^-$ &
$\left((\frac{-1}{6})^5 \,;\frac{1}{3}\,0\,
\frac{-1}{3}\,\right)
\left(\frac{1}{3}\,\frac{1}{3}\,\frac{1}{3}\,\frac{1}{3}\,0\,;
\frac{5}{12}\,\frac{-1}{4}\,
\frac{1}{2}\,
\right)'$ &  $\two'_{-5/3}$  & $1$ &$+4$&$0$ &$-4$ &$+8$ &$+5$ &$-5$ &$+234$\textbf{(+$\frac{13}{7}$)} & $D_{12}$& $+52$ \\[0.5em]
\hline\hline
& &  & &$-16$ &$-28$  &$+8$ &0&$+18$ &$+6$ &$-3492\qquad\quad$ &$\sum_i =$ & $-784$  \\[0.3em]
\hline

\end{tabular}
\end{center}
\caption{The SU(2)$'$ representations. Notations are the same as in Table \ref{tb:colorfields}. We listed only the upper component of SU(2)$'$ from which the lower component can be obtained by applying $T^-$ of SU(2)$'$.  }\label{tb:SU2fields}
}
\end{table}

There already exists an early attempt to obtain a CKM matrix from standard-like models implied by a $\Z_2\times\Z_2$ fermionic construction \cite{Faraggi94}.
Here, we attempt to obtain CKM and PMNS matrices based on the model of Ref. \cite{Huh09}. As commented before, if a high dimension superpotential term of light fields is generated then the relevant mass suppression factor must be the intermediate scale \Mit.  Suppose we have an effective operator for the $\Qem=\frac23$ quarks,
 \begin{eqnarray}
\frac{1}{M^n} {\rm (SM~singlets~of~
 Table~\ref{tb:singlets} )}\cdot \tenb_{-1}\five_3\five_{-2} .\label{eq:effMu}
 \end{eqnarray}
 
In Tables \ref{tb:SUhfields} and \ref{tb:SU2fields}, we list all the particles which transform non-trivially under SU(5)$'\times$SU(2)$'$.
There is no vector-like pair including  $\Qanom$ charges. Thus, there is no tree diagram of intermediate state with mass \Mvt, and any operator with sub-GUT scale fields must have mass suppression parameter \Mit. Thus, the suppression mass in Eq. (\ref{eq:effMu}) must be \Mit.

With this in mind,  let us discuss the Yukawa matrices and the fermion masses.  
The $\Qem=+\frac23$ quark mass matrix, consistent with \Uanom~symmetry, the orbifold selection rules and the the multiplicity 2 conditions of the Higgs doublets and matter fermions in $T_4^0$, is
  \begin{eqnarray}
  &\hskip -0.3cm{\five}_3^U\quad {\five}_3^{T_4^A} ~~{\five}_3^{T_4^S} &\nonumber \\[0.3em]
 M_u\propto \begin{array}{c} {\tenb}_{-1}^U\\[0.3em] 
  {\tenb}_{-1}^{T_4^A} \\[0.3em]   {\tenb}_{-1}^{T_4^S}\end{array} 
& \begin{pmatrix}\frac{\sigma_2\sigma_4}{\Mi^2},&0,& \frac{\alpha_1\sigma_4}{\Mi}\\[0.3em]
 0,&0,&1\\[0.3em]
 \frac{\alpha_2\sigma_2}{\Mi},& 1,&0
 \end{pmatrix} v_u&,~~{\rm with}~\langle H_u\rangle=\frac{v_u}{\sqrt2}.\label{eq:dmass}
  \end{eqnarray}
where superscripts $S$ and $A$ in $T_4^0$ denote the symmetric and antisymmetric combinations of  the multiplicity 2 fields, and $\alpha_1$ and $\alpha_2$ are coupling parameters. Similarly, the $\Qem=-\frac13$ quark mass matrix is given by
   \begin{eqnarray}
  &\hskip -0.2cm {\tenb}_{-1}^U~~{\tenb}_{-1}^{T_4^A} ~~{\tenb}_{-1}^{T_4^S} &\nonumber\\[0.3em]
 M_d\propto \begin{array}{c} {\tenb}_{-1}^U\\[0.3em] 
  {\tenb}_{-1}^{T_4^A} \\[0.3em]   {\tenb}_{-1}^{T_4^S}\end{array} 
& \begin{pmatrix}\frac{\sigma_4^2}{\Mi^2},&\quad 0~,& \quad \frac{\beta_1\sigma_4}{\Mi}\\[0.3em]
~ 0~~,&\quad 0~,&1\\[0.3em]
 \frac{\beta_1\sigma_4}{\Mi},&\quad 1~,&0
 \end{pmatrix} v_d&,~~{\rm with}~\langle H_d\rangle=\frac{v_d}{\sqrt2},\label{eq:umass}
  \end{eqnarray}
where $\beta_1$ is a coupling parameter. 
  
In Table \ref{tb:ChSinglets}, we listed charged singlets which will be needed for the charged lepton Yukawa couplings. Here, ${\one}_{-5}$ in $U$ and $T_4^0$ are $\Qem=+1$ charged leptons.  Thus, $\Qem=-1$ charged lepton mass matrix is 
   \begin{eqnarray}
  &\hskip -0.3cm{\five}_3^U\quad {\five}_3^{T_4^A} ~~{\five}_3^{T_4^S} &\nonumber \\[0.3em]
 M_e\propto \begin{array}{c} {\one}_{-5}^U\\[0.3em] 
  {\one}_{-5}^{T_4^A} \\[0.3em]   {\one}_{-5}^{T_4^S}\end{array} 
& \begin{pmatrix}\frac{\sigma_2\sigma_3}{\Mi^2},&0,& \frac{\gamma_1\sigma_3}{\Mi}\\[0.3em]
 0,&0,&1\\[0.3em]
 \frac{\gamma_2\sigma_2}{\Mi},& 1,&0
 \end{pmatrix}v_d&.\label{eq:emass}
  \end{eqnarray}
   
  Similarly, we can write the neutrino mass matrix, 
   \begin{eqnarray}
  &\hskip -0.5cm{\five}_{+3}^{U}\qquad {\five}_{+3}^{T_4^A} \qquad {\five}_{+3}^{T_4^S} &\nonumber \\[0.3em]
 M_\nu\propto \begin{array}{c} {\five}_{+3}^{U}\\[0.3em] 
 {\five}_{+3}^{T_4^A}\\[0.3em]  {\five}_{+3}^{T_4^S}\end{array} 
& \begin{pmatrix}\frac{\sigma_4\sigma_6^2\sigma_{13}\sigma_{16}}{\Mi^5},&0,& \frac{\{\sigma_2^2\sigma_4,\, \sigma_2\sigma_5\sigma_9\}}{\Mi^3}\\[0.3em]
 0,&0,&\frac{\sigma_1^3\sigma_4^2}{\Mi^5}\\[0.3em]
 \frac{\{\sigma_2^2\sigma_4,\, \sigma_2\sigma_5\sigma_9\}}{\Mi^3},& \frac{\sigma_1^3\sigma_4^2}{\Mi^5},&0
 \end{pmatrix}\frac{ v_u^2}{\Mi}.\label{eq:numass}
  \end{eqnarray}

The effective interaction for $M_\nu$ based on \smfa~introduces two $H_u$ insertions.  Note that $u$-type quarks obtain mass from the coupling $\tenb_{-1}\cdot\five_{+3}\cdot\five_{H_u}$ where $\five_{H_u}$ is $\five_{-2}$ in the flipped SU(5), and hence neutrino mass matrix comes from the square of this which introduces $v_u^2$. The coefficient of ${\five}_{+3}{\five}_{+3} v_u^2$ can be $\langle\tenb_{-1}\rangle^2/(\rm suppression~ mass)^3=\Mg^2/(\rm suppression~ mass)^3$ which we take as $1/\Mi$. In addition, there are nonzero factors of $\sigma^3/\Mi^3$ or $\sigma^5/\Mi^5$. For an illustration, take      $\sigma/\Mi=0.1-0.5$ and the largest neutrino mass $m_{\nu_{\rm max}}=0.5\,\eV$.  Then, $M_{\rm int}\approx 10^{8\,}-10^{11\,}\gev $. Obtaining $M_{\rm int}$ from the first principle is beyond the scope of this paper.

  Inspecting the mass matrices, we conclude that a physical phase in $\sigma_4$ leads to the CKM and PMNS phases.\footnote{The phases of other singlets can be rotated away by redefining the phases of some fermions.} For example, there can be a phase generated by the following superpotential
  \begin{eqnarray}
  W=m\sigma_6\sigma_8+\frac{1}{M^2}\sigma_6\sigma_7 \sigma_2\sigma_4^2
 \end{eqnarray}
where $m$ and $M$ are real parameters, and all fields develop nonvanishing VEVs. Then,
  \begin{eqnarray}
  \sigma_4=\pm i\,\left| m M^2\,\frac{\sigma_8}{\sigma_2\sigma_7}\right|^{1/2}e^{i(\delta_8-\delta_2-\delta_7)/2} 
 \end{eqnarray}
 where $\delta_i$ are the phases of $\sigma_i$. If $\delta_8=\delta_2=\delta_7=0$, the CKM and PMNS phases are determined as $\pm\frac{\pi}{2}$. 
 
The form of mass matrices in Eqs. (\ref{eq:dmass}),(\ref{eq:umass}),(\ref{eq:emass}), and (\ref{eq:numass}) can describe the quark and lepton masses successfully by various ratios of the singlet VEVs. Here, however, we will not try to find the relevant ratios.

 \subsection{The axion-photon-photon coupling}
  
Because we know all the $\Qanom$ charges of the electromagnetically charged fermions,  we can calculate the  axion-photon-photon coupling $\cagg$.  In Tables \ref{tb:SUhfields} and \ref{tb:SU2fields},  the typos in the previous tables \cite{KimPLB14} are corrected. Summing the $Q_a^{\gamma\gamma}$ columns of Tables \ref{tb:colorfields}, \ref{tb:ChSinglets}, \ref{tb:SUhfields}, and  \ref{tb:SU2fields}, we obtain
 \begin{equation}
 \cagg\simeq \frac{-9312}{-3492}-2=\frac{2}{3},
 \end{equation}
which must be the case if \Uanom~is the PQ symmetry \cite{KimNam16}. The subtraction of $\approx 2$, for $m_u/m_d\simeq 0.5$, is due to the contribution from the condensation of light quarks.
  
 \section{Conclusion}\label{sec:Concl}
We use the only allowed global symmetry \Uanom~from the heterotic string to find out the flavor structure of the SM. This global symmetry  is the most natural choice for the ``invisible'' axion from string theory. In addition,   \Uanom~is describing a flavor symmetry. In the  flipped SU(5) grand unification of Ref. \cite{Huh09}, we calculate mass matrices of quarks and leptons. It turns out that the fermion mass hierarchy in the SM results from the number of powers of Yukawa couplings, which is a common case in string compactification. Also, it is shown  how the doublet-triplet splitting in the flipped SU(5) GUT is realized in the model.
 
\acknowledgments

J.E.K. thanks J. Ashfaque and P.K.S. Vaudrevange for pointing out typos in Ref. \cite{KimPLB14}.
J.E.K. is supported in part by the National Research Foundation (NRF) grant funded by the Korean Government (MEST) (NRF-2015R1D1A1A01058449) and  the IBS (IBS-R017-D1-2016-a00), and B.K.  is supported in part by the   NRF-2016R1D1A1B03931151.
  
\section*{\bf Appendix: On the 't Hooft mechanism}

Superstring axions arising from the mother gauge symmetry $\EE8$ have been considered very early \cite{BarrAx85}. But, the case has been realized in \cite{KimPLB88,Lopez90,Halyo93} with the anomalous U(1) \cite{Anom87} after the orbifold compactification is known, which is possible even from the mother gauge symmetry SO(32) 
because the anomalous U(1) may arise from the spectra in the twisted sectors. Basically, Barr excluded SO(32) from the untwisted sector spectra \cite{BarrAx85}.  In all these cases, the working principle is the 't Hooft mechanism with the anomalous gauge U(1) from compactification where the U(1) is a subgroup of $\EE8$ or SO(32). In this string compactification, the Green-Schwarz (GS) counter term \cite{GS84} is essential.

The conventional 't Hooft mechanism in gauge theories \cite{Hooft71} is the following. Out of two continuous transformation parameters $\alpha(x)$ and $\beta$, acting on the field $\phi$,
\dis{
 \phi\to e^{i\alpha(x) Q_{\rm gauge}} e^{i\beta Q_{\rm global}}\phi,
}
 the $\alpha$ direction becomes the longitudinal mode of heavy gauge boson. The above transformation can be rewritten as
\dis{
 \phi\to e^{i(\alpha(x)+\beta) Q_{\rm gauge}} e^{i\beta (Q_{\rm global}-Q_{\rm gauge} )}\phi.
}
Redefining the local direction as $\alpha'(x)=\alpha(x)+\beta$, we obtain the transformation
\dis{
 \phi\to e^{i \alpha'(x) Q_{\rm gauge}} e^{i\beta (Q_{\rm global}-Q_{\rm gauge} )}\phi.
}
 So, the charge $Q_{\rm global}-Q_{\rm gauge}$ is reinterpreted as the new global charge and is not broken by the VEV, $\langle\phi\rangle$. Basically the direction $\beta$ remains as the unbroken continuous direction.
  
Consider the MI axion kinetic energy term in 10D \cite{GS84},
\dis{
-\frac{3\kappa^2}{2 g^4\,\varphi^2}H_{MNP}H^{MNP},~{\rm with}~{M,N,P}=\{1,2,\cdots,10\}  
}
which is parametrized after compactification to 4D as 
\dis{
\frac{1}{2\cdot 3! M_{MI}^2}H_{\mu\nu\rho}H^{\mu\nu\rho},~{\rm with}~{\mu,\nu,\rho}=\{1,2,3,4\}. \label{eq:H2term}
}
The GS action in the differential form is \cite{GS84}
\dis{
S_1'=\frac{c}{108\,000}\int\left\{ 30B\left[ ({\rm tr}_1 F^2)^2+( {\rm tr}_2 F^2)^2- {\rm tr}_1 F^2\,{\rm tr}_2F^2 \right]+ \cdots\right\} 
}
where ${\rm tr}_1$ and ${\rm tr}_2$ are for the E$_8$ and E$_8'$ representations, which is relevant for our model discussed in the paper.  In 4D, it leads to an interaction 
\dis{
S_1'\propto -\frac{c}{10800}\left\{
 H_{\mu\nu\rho}A_\sigma\,\epsilon^{\mu\nu\rho\sigma}\epsilon^{ijklmn}\langle F_{ij}\rangle \langle F_{kl}\rangle \langle F_{mn}\rangle  + \cdots\right\}\to \frac{1}{3!}\epsilon_{\mu\nu\rho\sigma}  H^{\mu\nu\rho} A^{\sigma}\label{eq:GSterm}
}
where we have taken VEVs of internal gauge field strengths, $F_{ij}$ etc. Superstring carries the electric field $A_\mu$, whose group space value is simply denoted as $A$, a kind of matrix. 

 \begin{figure}[!h]
\begin{center}
\includegraphics[width=0.4\textwidth]{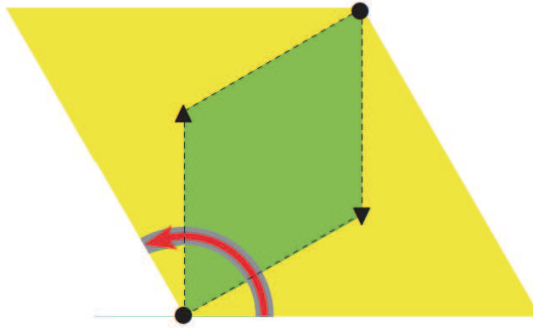} 
\end{center}
\caption{The flux $\langle F_{ij}\rangle$ at a fixed point.} \label{fig:Flux}
\end{figure}

First, let us note how we assign VEVs to field strengths in Eq. (\ref{eq:GSterm}). Then, we discuss the  't Hooft mechanism. The line integral around a fixed point gives an integer. But, in the orbifold compactification, we need to line-integrate only a part of $2\pi$. In Fig. \ref{fig:Flux}, we showed the (limegreen) fundamental domain of $Z_3$. The line integral is for matrix value gauge field $A_\mu$ which is in our case denoted by the shift vector and Wilson lines, $V,V\pm a_i$ in the $(ij)$ internal plane. By Stoke's theorem, it is $\frac13$ of the surface integral of $F_{ij}$ which is assumed to be located at the fixed point $F_{ij}\delta^2(\sigma-\sigma_0)$ where $\sigma_0$ is the location of the fixed point in the two-torus. This delta function limit of zero-length string must be considered for the string tension contribution to be neglected. In this way,  we assign VEVs to field strengths in Eq. (\ref{eq:GSterm}).

\begin{figure}[!t]
\begin{center}
\includegraphics[width=0.6\textwidth]{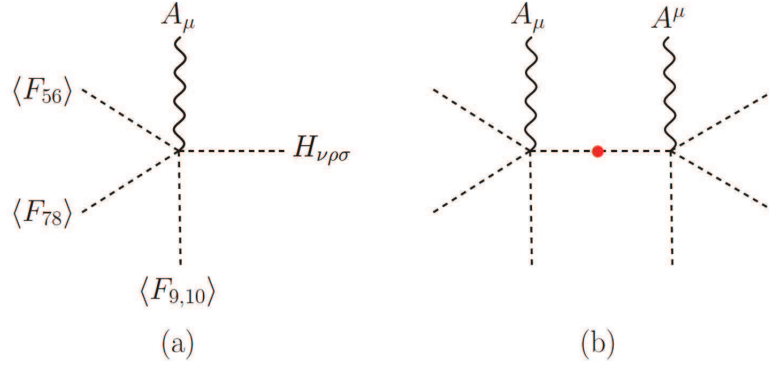} 
\end{center}
\caption{The GS terms.} \label{fig:GS}
\end{figure}

We note that one scalar degree $H_{\mu\nu\rho}$ is the MI-axion \cite{Witten84}
\dis{
 H_{\mu\nu\rho}=M_{MI}\epsilon_{\mu\nu\rho\sigma}\,\partial^\sigma a_{MI}.  \label{eq:afromH}
}
In Fig. \ref{fig:GS}\,(a), we show the first term of Eq. (\ref{eq:GSterm}), transferring one derivative of $F_{\mu\nu}$ to $B_{\rho\sigma}$.
Eqs. (\ref{eq:H2term}) and (\ref{eq:afromH}) give the following  
\dis{
\frac12 \partial^\mu  a_{MI}\partial_\mu a_{MI}+    M_{MI}A_\mu\partial^\mu a_{MI}.  \label{eq:HooftMI}
}
 The global direction  with respect to the 't Hooft mechanism is the $a_{MI}$ direction. 
Eq.  (\ref{eq:HooftMI}) can be expressed as, by adding the contribution of Fig. \ref{fig:GS}\,(b) [which is obtained from  Fig. \ref{fig:GS}\,(a)],
\dis{
\frac12  (\partial_\mu a_{MI})^2+  M_{MI}A_\mu\partial^\mu a_{MI}+\frac{1}{2}M_{MI}^2A_\mu A^\mu
 \to  \frac12 M_{MI}^2( A_\mu +\frac{1}{M_{MI}}\partial_\mu a_{MI})^2 . \label{eq:AmassGS}
}
Thus, the MI-axion degree is completely removed below the gauge boson mass scale where the heavy anomalous gauge boson including the longitudinal degree is defined as $\tilde{A}_\mu =A_\mu+(1/M_{MI})\partial_\mu a_{MI}$. The MI-axion is dynamicaly removed from the theory except that it couples to the anomaly \cite{Svrcek06}. This coupling is defining a $\theta$ parameter at low energy depending on some cosmological determination of $\langle a_{\rm MI}\rangle$.
The resulting global symmetry is broken by the usual Higgs mechanism and we expected that it is achieved at the intermediate scale. 

Now, let us denote the charge of U(1)$_a$ subgroup of $\EE8$ as   $Q_a$, which was anomalous in the beginning. Suppose that $\phi$ carries nonzero gauge charge $Q_a$. All terms involving $\phi$ should respect the original U(1)$_a$ gauge symmetry. So, the potential $V$ is a function of $\phi^*\phi$ only. 

Below $M_{MI}$, we do not consider the FI term   because there is no U(1)$_a$ gauge symmetry. But, to discuss the scales of the  U(1)$_{\rm anom}$ breaking, we consider  the $\phi$ couplings together with the $a_{MI}$ degree. The $\phi$ coupling to $A_\mu$ is given by, for $\phi=(\frac{v+\rho}{\sqrt2})e^{ia_{\phi}/v}$,
\dis{
 |D_\mu \phi|^2 &=|(\partial_\mu -ig
Q_{a}A_\mu)\phi|^2_{\rho=0}= \frac12(\partial_\mu a_{\phi})^2-gQ_a A_\mu \partial^\mu a_{\phi}+\frac{g^2}{2}Q_a^2v^2 A_\mu^2 \\
 &=\frac{g^2}{2}Q_a^2v^2( A_\mu-\frac{1}{gQ_a v}\partial^\mu a_{\phi})^2. \label{eq:AmassPhi}
}
The gauge boson $A_\mu$ obtains mass from two contributions, Eqs. (\ref{eq:AmassGS}) and (\ref{eq:AmassPhi}), which can be written as
\dis{
 \frac12\left(M_{MI}^2+g^2Q_a^2v^2 \right)(A_\mu)^2+A_\mu( M_{MI}\partial^\mu a_{MI} -g Q_a v \partial^\mu  a_{\phi} )+\frac12\left[ (\partial_\mu a_{MI})^2+(\partial^\mu  a_{\phi})^2 \right]   \qquad \qquad  (9)
}
Let us define a new global degree as $a$ interpretable as the ``invisible'' axion,
\dis{
a=\cos\theta\,a_{\phi} +\sin\theta\,a_{MI} }
where
\dis{
 \sin\theta=\frac{g Q_a v}{\sqrt{M_{MI}^2+g^2Q_a^2v^2}}.  
}
Thus, if $v\ll M_{MI}$,  the global symmetry breaking scale can be at the intermediate scale and the axion has the dominant component from the phase of $\phi$. Determination of $v$  in the effective field theory framework is such that the coefficient of $\phi^*\phi$ in the potential is given by $-(\rm intermediate~ scale)^2$.


\end{document}